\documentclass[aps,prd,preprint,superscriptaddress,showpacs,nofootinbib,showkeys]{revtex4-1}

\usepackage{amsmath}
\usepackage{graphicx}

\usepackage{hyperref} 

\usepackage{color}

\begin{document}

\title{Dark matter perturbations and viscosity: a causal approach}

\author{Giovanni Acquaviva}
\email[]{gioacqua@utf.troja.mff.cuni.cz} \affiliation{Institute of Theoretical Physics, Faculty of
Mathematics and Physics, Charles University in Prague, 18000 Prague, Czech Republic}

\author{Anslyn John}
\email[]{{ajohn@sun.ac.za}}
\affiliation{National Institute for Theoretical Physics (NITheP), Stellenbosch 7600, South Africa}
\affiliation{Institute of Theoretical Physics, Stellenbosch University, Stellenbosch 7600, South Africa}

\author{Aur\'{e}lie P\'{e}nin}
\email[]{aurelie.c.penin@gmail.com}
\affiliation{Astrophysics and Cosmology Research Unit, School of Mathematical Sciences, University of KwaZulu-Natal, Durban 4041, South Africa }

\date{\today}

\begin{abstract}

The inclusion of dissipative effects in cosmic fluids modifies their clustering properties and could have observable effects on the formation of large scale structures.  We analyse the evolution of density perturbations of cold dark matter endowed with causal bulk viscosity.  The perturbative analysis is carried out in the Newtonian approximation and the bulk viscosity is described by the causal Israel-Stewart (IS) theory. In contrast to the non-causal Eckart theory, we obtain a third order evolution equation for the density contrast that depends on three free parameters.  For certain parameter values, the density contrast and growth factor in IS mimic their behaviour in $\Lambda$CDM when $z\geq 1$. Interestingly, and contrary to intuition, certain sets of parameters lead to an increase of the clustering.

\end{abstract}


\maketitle

\section{\label{intro}Introduction}

The $\Lambda$CDM model is the simplest and most coherent description of the background evolution of
the observed universe, from the cosmic microwave background (CMB) epoch to the present phase of accelerated expansion.  While this framework is supported by many cosmological observations -- the CMB anisotropies and Supernovae Ia amongst others \cite{Planck2015,riess20162} -- several inconsistencies remain between the $\Lambda$CDM dynamics of structure formation and observations. For instance, the `missing satellite problem' where N-body simulations predict too many satellite galaxies as compared to those observed around the Milky Way \cite{Moore1999} or the `core-cusp problem' where the same simulations produce  halo density profiles more cuspy than those measured in the center of dwarf galaxies \cite{Donato2009}. Both issues imply that pressureless or cold dark matter (CDM) produces an excess of structure and clustering compared to what we observe.

Several solutions within the CDM framework have been considered, such as the inclusion of baryonic feedback, which is a possible way of solving both issues \cite{Brooks2013,Zolotov2012}. Many alternatives to simple CDM have also been put forward ranging from warm dark matter to more radical modifications of the theory of gravitation \cite{Moffat2006,Capozziello2007,BOHMER2008}. One possibility 
is the modification of the properties of the fluids or fields \cite{Bertone2005,Li2009}.  A minimal extension would
entail the relaxation of the hypothesis of exact equilibrium in the description of the CDM fluid: this would
amount to the inclusion of dissipative effects in the cosmic fluid and, as a consequence, a deviation from
its pressureless character specified by the equation of state parameter $w=0$. The main consequence is a suppression of small scale structures as compared to the pure CDM scenario. In addition, bulk
viscosity has another important feature for cosmology: it can generate an accelerated expansion era
without invoking dark energy.\footnote{However, one has to keep in mind that the assumptions
underlying the description of cosmic viscous fluids could break down during inflation, as the latter
represents a strongly out-of-equilibrium scenario where the hydrodynamic framework could be
unreliable.}\\

The inclusion of viscosity in the cosmic fluid and its effect on the growth of large-scale structures has
been considered by many authors.  In \cite{Li2009} it is shown that, when describing the entire dark sector by a single viscous fluid, the dynamic of perturbations is poorly reconciled with observations; in \cite{Velten2013} was shown that the Newtonian description of viscous matter clustering is unreliable and that at least a neo-Newtonian treatment is necessary. 
However, most of the works regarding viscous cold dark matter make use of Eckart theory \cite{Eckart1940}, which is a non-causal approach to dissipative phenomena. Therefore, relaxation to equilibrium is considered as instantaneous, leading to an infinite propagation speed of the density perturbations in the fluid. In this paper, we extend a previous analysis by including a non-vanishing relaxation time, $\tau$, in the transport equation for bulk viscosity following the framework introduced by Israel and Stewart (IS) \cite{Israel1979}.  An analysis of the gravitational potential in a cosmological context taking into account causal dissipative processes has already been presented in \cite{1475-7516-2011-05-029}, where it was shown that the truncated version of IS is favoured over both Eckart and the full IS.  However, they adopt an ansatz for the functional form of the viscous pressure while in the present analysis we implement the full transport equations in order to determine the evolution of the viscous pressure and its perturbation.\\

The paper is structured as follows: in section \ref{theory} we develop the theoretical framework, describing
the background dynamical equations and deriving the evolution equation of density perturbations for 
viscous CDM in the IS framework; in section \ref{qualitative} we perform a qualitative analysis of the
evolution equation focusing on the time-scale $\tau$ introduced by IS theory; in section \ref{results} we present  numerical solutions of the evolution equation and comment on their properties. We draw our concluding remarks in section \ref{conclusions}.  Unless otherwise specified we use units in which $c=1$.

\section{\label{theory}Theoretical setup}

\subsection{Background dynamics}

Since we are interested in the growth of structures on sub-horizon scales, we use the Newtonian theory
of gravity as our starting model. This is a reliable approximation to general relativity when describing
non-relativistic matter on scales well within the Hubble radius, {\it e.g.} at galaxy cluster or filament scales.
The evolution and propagation of non-relativistic matter in Newtonian cosmology is described by the following
system:
\begin{eqnarray}
\frac{\partial \rho}{\partial t} + \nabla \cdot \left( \rho \mathbf{u} \right) &=& 0 \label{navier1} \\
\rho \left( \frac{\partial}{\partial t} + \mathbf{u} \cdot \nabla  \right) \mathbf{u} = - \nabla p - \nabla \Pi &-&  \rho \nabla \Phi \label{navier2} \\
\nabla^2 \Phi &=& 4 \pi G \rho\ \label{navier3},
\end{eqnarray}
where $\rho$ is the mass density, $\bf{u}$ is the fluid velocity, $\Phi$ is the gravitational potential and
the partial derivative is taken with respect to the cosmic time. The total fluid pressure has been
separated into equilibrium ($p$) and dissipative ($\Pi$) contributions. The bulk viscous pressure $\Pi$ satisfies
a transport equation given by IS causal theory of dissipation:
\begin{equation}\label{IS}
 \tau \dot{\Pi} + \Pi = - \zeta\, \nabla \cdot \mathbf{u} - \frac{\epsilon}{2}\, \Pi\, \tau \left[ \nabla \cdot \mathbf{u} + \frac{\dot{\tau}}{\tau} - \frac{\dot{\zeta}}{\zeta} - \frac{\dot{T}}{T} \right]\ .
\end{equation}
Hereafter the overdot represents the time derivative in comoving coordinates, which in the Newtonian
approximation is just the convective derivative, $\frac{D}{Dt}=\frac{\partial}{\partial t}+ \bf{u}\cdot
\nabla$. The full IS theory is obtained when the bookkeeping parameter $\epsilon=1$, while a truncated version (TIS) is given by
$\epsilon=0$. The non-causal Eckart theory is recovered when the characteristic relaxation time
$\tau\rightarrow 0$. $T$ is the fluid temperature.

The quantity $\zeta$ is the coefficient of bulk viscosity, which is taken to be
of the form
\begin{equation}\label{zeta}
\zeta = \zeta_0 \left( \frac{\rho}{\rho_0} \right)^s\ ,
\end{equation}
where $s$ is a constant and the energy density ($\rho c^2 \equiv \rho $) evaluated at the present time ($a_0=1$) is denoted by $\rho_0$. The present value of the bulk viscosity is $\zeta_0$.  If one considers a cosmological scenario with a single barotropic fluid component having a linear equation of state $p = w\, \rho$, the inclusion of bulk viscosity leads to a total effective pressure $p_{eff}=w\, \rho + \Pi$.  In the Eckart theory this is equivalent to considering a fluid with a nonlinear equation of state $p_{eff}= w\, \rho - w_b\, \rho^{s+1/2}$, where $w_b\geq0$.  Such a straightforward interpretation is not applicable in the presence of additional cosmological fluids and it doesn't hold for the extension to the IS framework. Indeed, in the latter case the relation between $\Pi$ and the energy density is not algebraic anymore but is mediated by the evolution equation Eq.(\ref{IS}).

The relaxation time $\tau$ is defined in terms of the sound speed of bulk viscous perturbations \cite{maartens1996} via 
\begin{equation}
c_{b}^{2} = \frac{\zeta}{\left( \rho + p \right) \tau}.  
\end{equation}
If the matter component obeys the linear equation of state $p = w \rho$ then the relaxation time becomes
\begin{equation}\label{tau}
\tau = \frac{\zeta_0\, \rho^{s-1}}{\left( 1 + w \right)\, c_{b}^{2}\, \rho_0^s}\ .
\end{equation}
The dissipative sound speed contributes together with the adiabatic sound speed $c_s^2$ to form the total sound speed $v^2=c_b^2+c_s^2$. For dust $c_s^2=0$, therefore causality requires $c_b$ to be less than the speed of light $c$. Moreover, a finite degree of clustering requires $c_b^2\ll c^2$ because relativistic particles cannot form structures.  Bounds on the value of the adiabatic speed of sound for perfect fluids have been inferred from galaxy cluster mass profiles in \cite{sartoris2014clash} and from the observed rotation curves of spiral galaxies in
\cite{avelino2015constraints}.  In the absence of analogous observational constraints on dissipative effects, we consider those results as bounds on the {\it total} sound speed $v$ and hence on the dissipative part $c_b$. We will thus employ the most conservative constraint $c_b^2 < 10^{-8} c^2$.

Finally, assuming a linear equation of state for the fluid component, the integrability condition of the Gibbs
relation leads to 
\begin{equation}\label{temp}
T=T_0\, \rho^{\frac{w}{1+w}}\ .
\end{equation}
Hence, once the equation of state of the viscous fluid is specified, the free parameters in IS theory are $c_b^2$, $s$ and $\zeta_0$.  It will be convenient to express the latter in the dimensionless combination
\begin{equation}
 \tilde{\zeta} = \frac{24\, \pi\, G}{H_0}\, \zeta_0\, .
\end{equation}

\subsection{Perturbed equations}

We now derive the general evolution equation for viscous CDM perturbations.  First of all, the energy density
of the fluid can be split into background $\rho$ and first order perturbation $\delta\rho$.  In the following
we will focus on the evolution of the density contrast $\delta\equiv\delta \rho/\rho$.  Perturbing and
linearising (\ref{navier1}) - (\ref{navier3}) yields the following system:
\begin{eqnarray}
\dot{\delta} + \frac{1}{a} \nabla \cdot \mathbf{v} &=& 0 \\
\dot{\mathbf{v}} + H \mathbf{v} = - \frac{1}{a \rho} \nabla \left( \delta p \right) -  \frac{1}{a \rho} \nabla \left( \delta \Pi \right) &-& \frac{1}{a} \nabla \left( \delta \Phi \right) \\
\nabla^2 \left( \delta \Phi \right) &=& 4 \pi G a^2 \rho \delta.
\end{eqnarray}
Combining these equations and using a linear equation of state, we can write down the general
result, valid for any pressure source,
\begin{equation}\label{ddot}
 \ddot{\delta} + 2H\, \dot{\delta} - 4\pi G\, \rho\, \delta = -\frac{w}{a^2} k^2\, \delta - \frac{1}{a^2 \rho} k^2\, \delta\Pi
\end{equation}
In this last expression we have already performed the substitution $\nabla\rightarrow i k$, which is the result of
a spatial Fourier transform of the perturbations.  The bulk viscosity and the relaxation time can be expanded
to first order in $\delta$ as
\begin{align}
\zeta + \delta \zeta &\simeq \zeta + s\, \zeta\, \delta \label{dzeta}\\
\tau + \delta \tau &\simeq \tau + (s-1)\, \tau\, \delta \label{dtau}\, .
\end{align}
These expressions are the basis of the derivation of the general evolution equation for the density perturbations as a function of the scale factor viz. 
\begin{align}\label{ppprime}
 H \tau\, a^3\delta''' &+ \Big\{ \Big[ 3(\epsilon - q) + 1 \Big] H \tau + 1 \Big\}\, a^2\, \delta'' \nonumber\\
 &+ \left\{ \left[ \left( 3\epsilon (2-q) + j - 3 q - 4 \right) - \frac{4\pi G \rho}{H^2} + \epsilon \frac{k^2\, \Pi}{a^2H^2\rho}  \right] H \tau + (2-q) + \frac{k^2 \zeta}{a^2H\rho} \right\}\, a\, \delta'\nonumber\\
 &+\left\{ \left[ \frac{4\pi G \rho}{H^2}(4-3\epsilon) \right] H \tau + \frac{k^2}{a^2H^2 \rho} \Big( (s-1) \Pi - 3H\zeta \Big) - \frac{4\pi G \rho}{H^2} \right\}\, \delta = 0\ ,
\end{align}
where the prime denotes derivatives with respect to the scale factor.  We would like to stress that in the limit $\tau \rightarrow 0$ (or equivalently $c_b^2\rightarrow\infty$) Eq.\eqref{ppprime} correctly reduces to Eckart's form \cite{Velten2013}.  The non-viscous $\Lambda$CDM case is then recovered for $\zeta_0\rightarrow 0$. The derivation of Eq.\eqref{ppprime} is detailed in appendix \ref{deriv}.\\

The deceleration, $q=-\ddot{a}\, a\, \dot{a}^{-2}$, and jerk, $j=-\dddot{a}\, a^2\, \dot{a}^{-3}$, parameters can be written in terms of $a$, $H$ and its derivatives:
\begin{align}
 q &= -1-a\, \frac{H'}{H}\\
 j &= -1+a^2 \frac{H'^2}{H^2}+a \left(4 \frac{H'}{H}+a \frac{H''}{H}\right)
\end{align}
For pressureless dust, the background energy density is $\rho=\rho_0\,
a^{-3}$. Defining the fractional energy density at the present time $\Omega_{0}=\frac{8\pi G}{3}\, \rho_{0}\, H_0^{-2}$, the background energy density can be rewritten as 
\begin{equation}
 \rho(a) = \frac{3 H_0^2\, \Omega_0}{8\pi G}\ a^{-3}. \label{rhoa}
\end{equation}
In general, the functional dependence of $H$ on the scale factor depends on the species present in
the background.  For baryonic and dark matter, radiation and a cosmological constant, Friedmann equation gives
\begin{equation}
 H^2(a) = H_0^2\, \left[ \left( \Omega_{0} + \Omega_{b0} \right)\, a^{-3} + \Omega_{r0}\, a^{-4} + \Omega_{\Lambda} \right] \label{hubblea}
\end{equation}
where $\Omega_{b0}$ and $\Omega_{r0}$ are the fractional densities of baryons and radiation respectively,
evaluated at present time, while $\Omega_{\Lambda}$ is the cosmological constant contribution.  The
parameters that are given by observations are $\{ H_0 , \Omega_{0} , \Omega_{b0}, \Omega_{\Lambda} \}$.  For the purpose of our analysis we will disregard the baryonic sector, since its contribution is negligible compared to the dark matter one.\\

Finally, the background function $\Pi$ in Eq.\eqref{ppprime} is determined by solving numerically Eq.\eqref{IS} with the boundary condition that the bulk viscous pressure satisfies Eckart's relation at some initial time $a_i$, {\it i.e.} $\Pi(a_i)=-3\, H(a_i)\, \zeta(a_i)$.\\

Before presenting the results of our analysis, we would like to stress that the setup employed here, expressed by Eq.\eqref{ppprime} coupled with the background quantities $\Pi$ and $H$ given by Eq.\eqref{IS} and Eq.\eqref{hubblea} respectively, relies on a framework in which Newtonian perturbations evolve in a relativistic background. Such choice is motivated by the range of scales we are interested in, {\it i.e.} subhorizon scales in a linear regime, where the Newtonian and relativistic treatments rapidly converge.  We also stress the fact that the inclusion of relativistic species at background level -- as it is done in eq.\eqref{hubblea} -- does not affect the validity of the Newtonian approximation at perturbative level, as long as the conditions for the latter are met.

\section{\label{qualitative}Qualitative analysis}
In this section we develop qualitative arguments describing the impact of free parameters on the evolution of the density contrast given by  Eq.\eqref{ppprime}.  There are three characteristic timescales appearing in the problem: (i) the expansion time $t_e\sim H^{-1}$, (ii) the collapse time $t_c=(4 \pi G\, \rho)^{-1/2}$ and (iii) the relaxation time $\tau=(\zeta_0\, \rho^{s-1})/(c_b^2\, \rho_0^s)$. As we consider only the matter-dominated phase of expansion, $t_e=2/(3H)$.  Making use of the Friedmann equation, it is clear that any ratio between $t_e$ and $t_c$ is constant and only ratios involving $\tau$ change over time. In Eq.\eqref{ppprime} the relaxation time appears exclusively in ratios with the form:
\begin{equation}
 H\, \tau = \frac{2}{3}\frac{\tau}{t_e}
\end{equation}
We will thus consider $\tau/t_e$ as a measure of the deviation between IS and Eckart. In the limit
$\tau/t_e\rightarrow 0$ the two theories coincide while if $\tau/t_e\gtrsim1$ the terms involving the
relaxation time are, in general, not negligible and the two theories differ.  Remembering that in the matter era we have $\rho=\rho_0\, a^{-3}$, we express the condition for negligible departure from Eckart as
\begin{equation}\label{condition}
 \frac{\tau}{t_e}\ll 1\ \ \ \ \Rightarrow\ \ \ \ \frac{\zeta_0}{c_b^2}\sqrt{\frac{8 \pi G}{3 \rho_0}}\ a^{3\left(\frac{1}{2}-s\right)}\ll 1
\end{equation}
It is immediately apparent that this condition depends on the ratio of the two parameters $\zeta_0$ and
$c_b^2$. However, once the values of such time-independent parameters are fixed, there is a more crucial
dependence on $s$ in the exponent of the scale factor.  In particular {\it the choice of the exponent,
irrespective of the other parameters, affects the deviation between Eckart and IS in the following way:}
\begin{itemize}
  \item if $s<1/2$ then the condition $\tau/t_e\ll 1$ always holds for $a\ll 1$: IS and Eckart coincide at
      early times and the deviations can show up only at later times;
  \item if $s>1/2$ then the opposite condition $\tau/t_e\gg 1$ holds at early times, meaning that a
      significant deviation between IS and Eckart already appears for $a\ll 1$.
\end{itemize}
In both cases the degree of the deviation is governed by the specific value of the amplitude
$\frac{\zeta_0}{c_b^2} \sqrt{\frac{8 \pi G}{3 \rho_0}}$.  It is worth stressing that the importance of the
value $s=1/2$ has been extensively reported in previous studies regarding the background evolution of
cosmological models: in \cite{barrow1988string,acquaviva2015nonlinear}, for example, it was shown that an accelerated expansion of the scale factor can be obtained at late (early) times if $s<1/2$ ($s>1/2$).  The present analysis shows that $s=1/2$ also has a role at the perturbative level and represents a dividing value between early and late-time features of the density contrast growth.  In the next section we present numerical results that clearly illustrate this feature. 

\section{\label{results}Results}
In this section, we solve Eq.\eqref{ppprime} numerically. We set initial conditions at the time of decoupling ($z\sim1100$), when matter starts dominating the total energy density while radiation becomes negligible. We use CAMB to compute the linear power spectrum of matter \cite{lewis2000efficient} with Planck cosmological parameters \cite{Planck2015}  to compute the initial conditions.  As we focus mostly on the matter-dominated epoch, we neglect the contribution of baryons.\\

\begin{figure*}[th!!!]
  \begin{center}
    \includegraphics[width=0.47\textwidth]{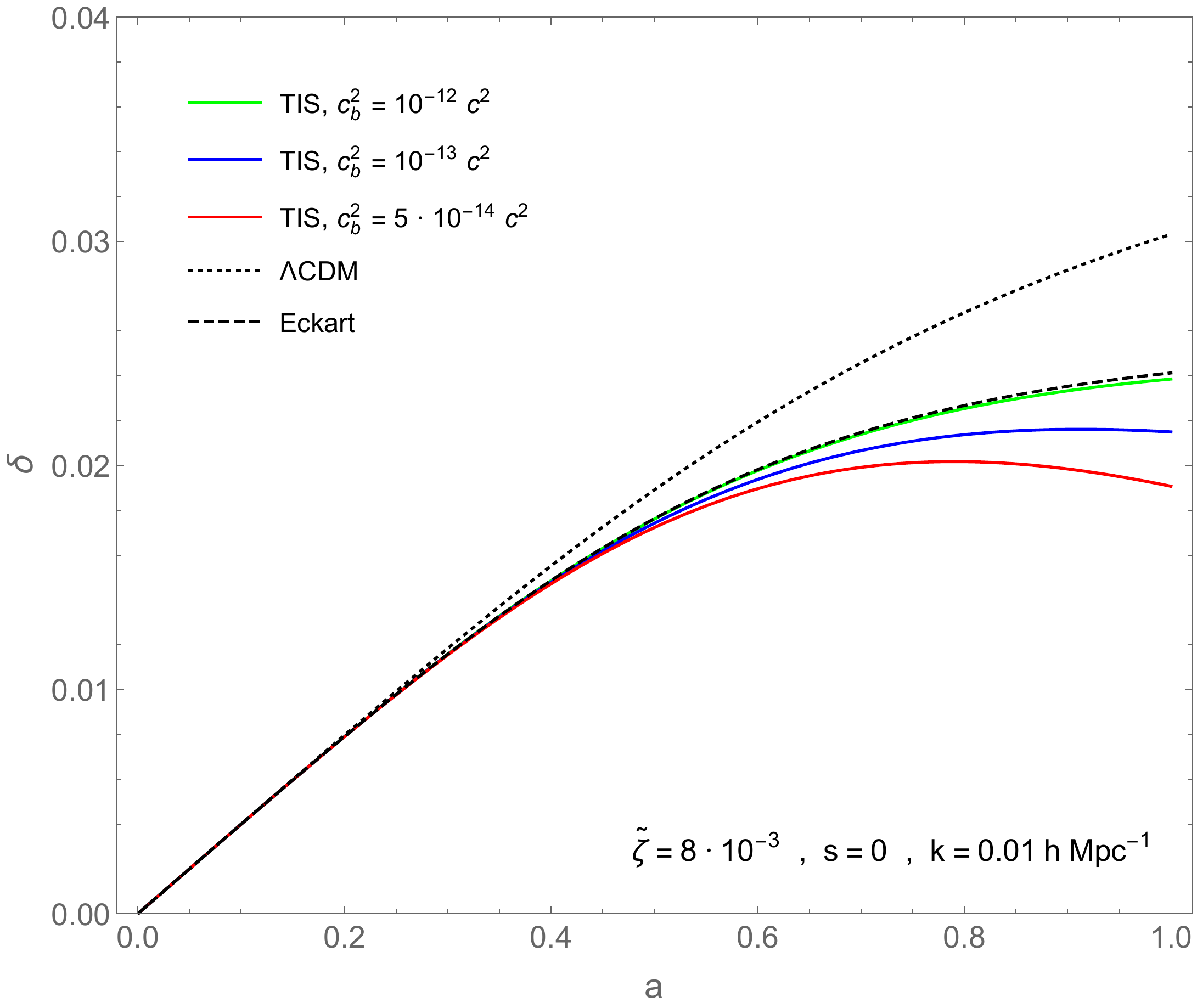}
    \includegraphics[width=0.47\textwidth]{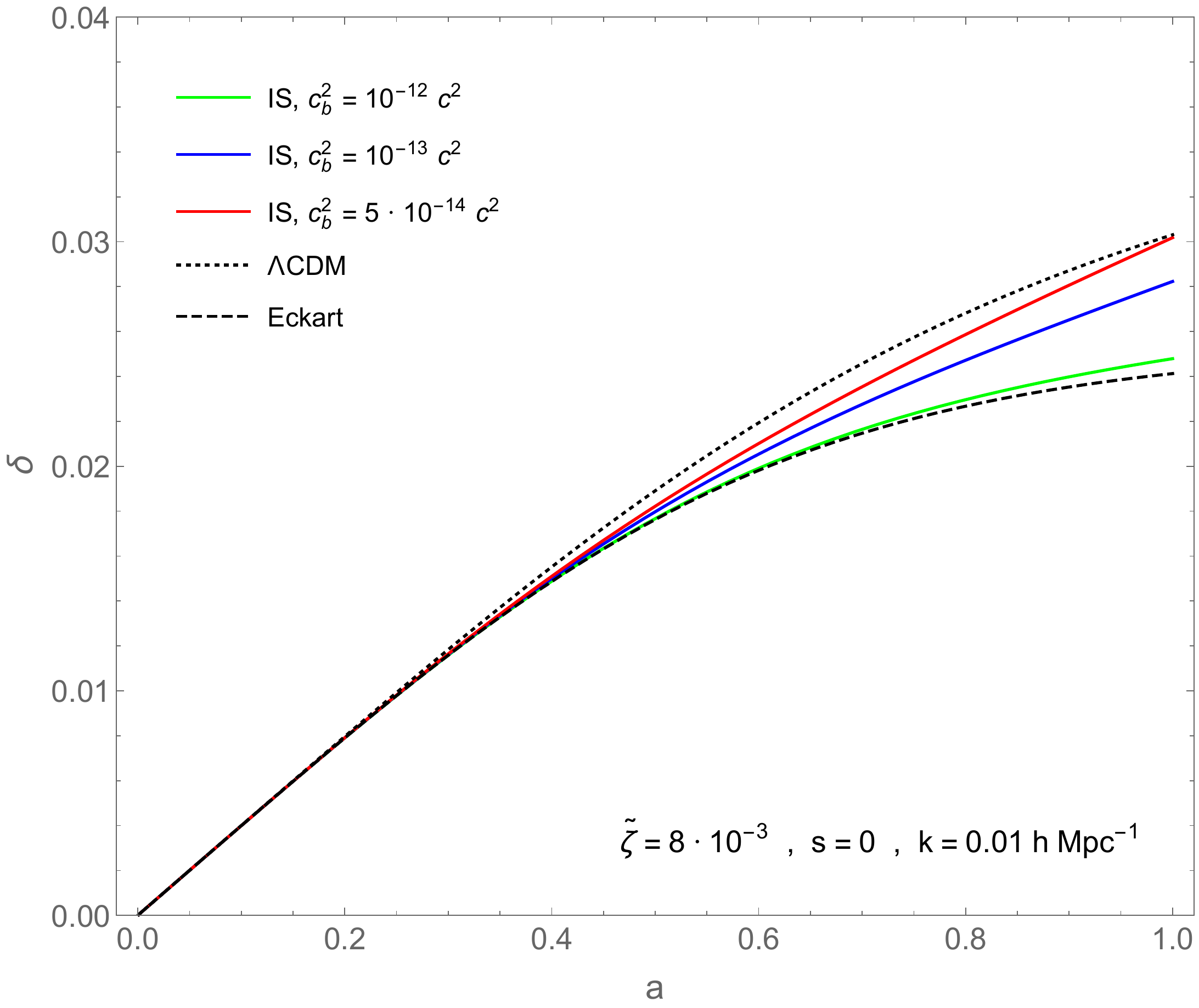}
    \includegraphics[width=0.455\textwidth]{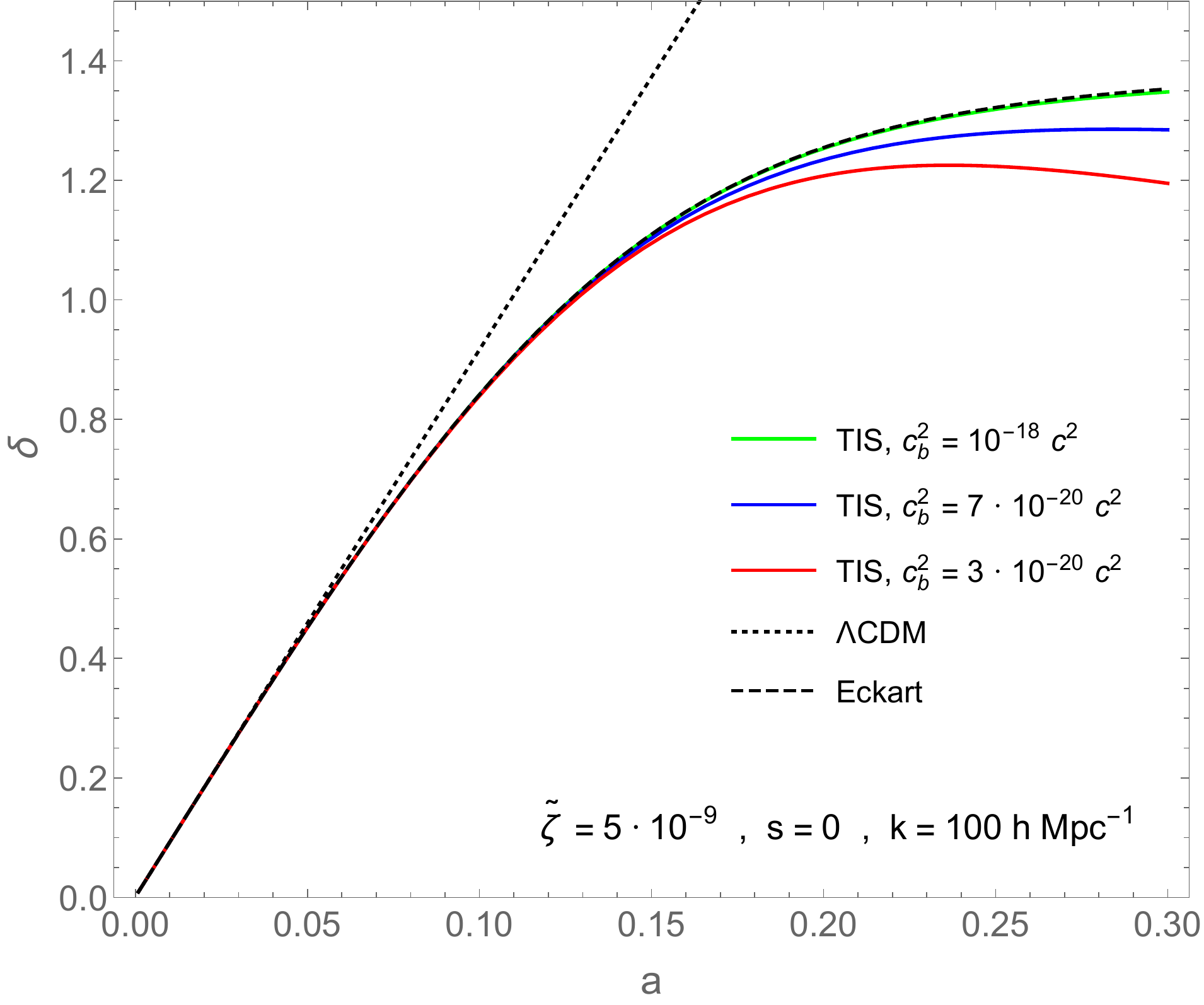}
    \includegraphics[width=0.455\textwidth]{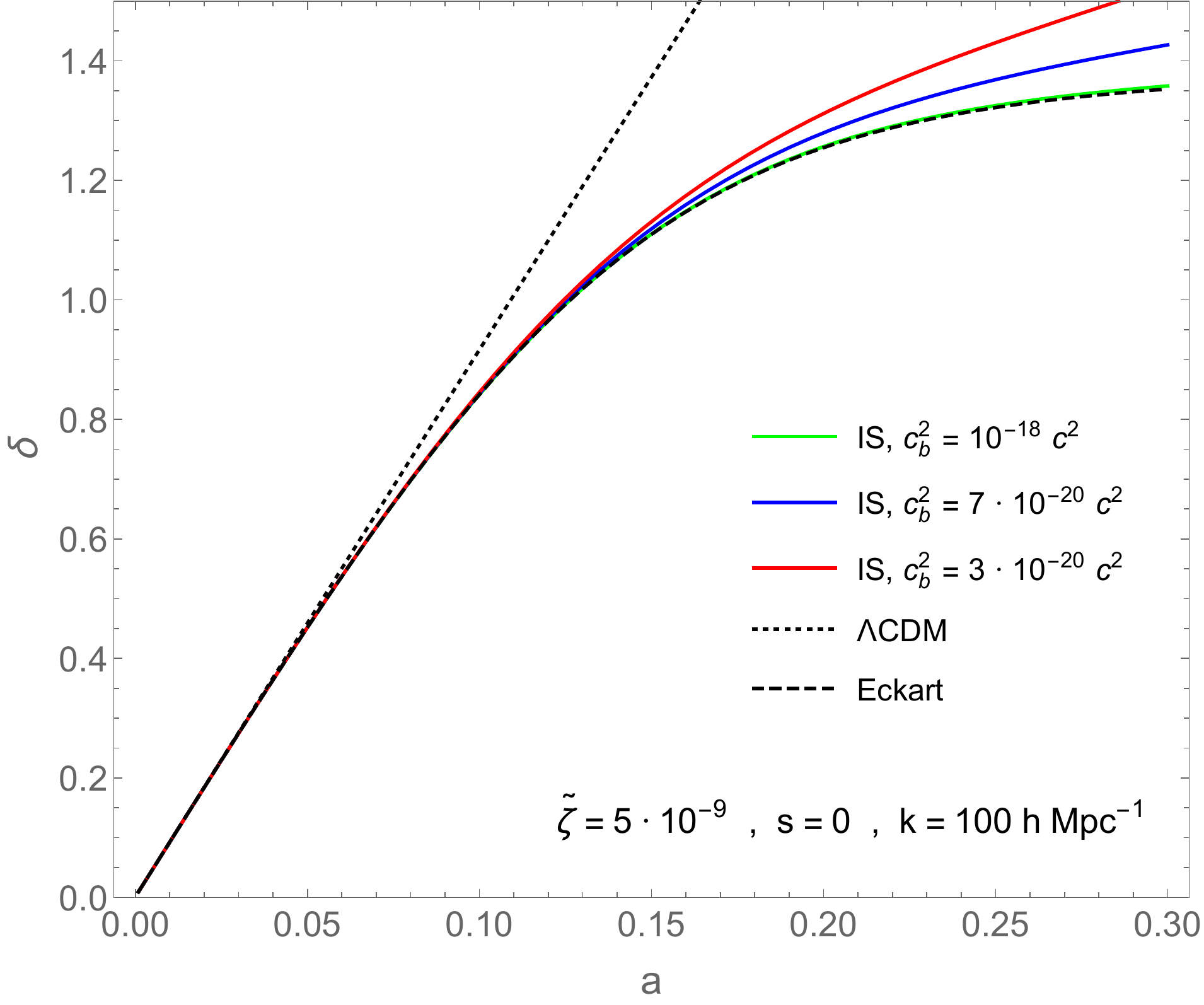}
  \caption{\label{delta_s0} Evolution of density contrast for $s=0$ varying $c_b^2$ in TIS (left panels) and IS (right panels), for $k=0.01 \ h$ Mpc$^{-1}$ (top panels) and $k=100\ h$ Mpc$^{-1}$ (bottom panels).}
  \end{center}
\end{figure*}

First, we investigate the effect of $c_b^2$ on the evolution of the density contrast.  In Fig.\ref{delta_s0} we plot $\delta(a)$ for constant bulk viscosity ($s=0$) at two different scales, $k=0.01\, h$ Mpc$^{-1}$ and $k=100\, h$ Mpc$^{-1}$ (galaxy scale), for a fixed value of $\tilde{\zeta}$ and three values of $c_b^2$.  The values of the parameters chosen show a significant deviation both from Eckart's curve and from the standard $\Lambda$CDM result.  It is worth stressing that deviations that occur near and beyond $\delta\simeq1$ are not reliable, because that represents the threshold of the linear regime of perturbations.  Nonetheless, one can appreciate the qualitative difference between TIS and IS: while the former leads to a further suppression with respect to Eckart's result, the latter enhances the density contrast at late times.

\begin{figure*}[th!!!]
  \begin{center}
    \includegraphics[width=0.47\textwidth]{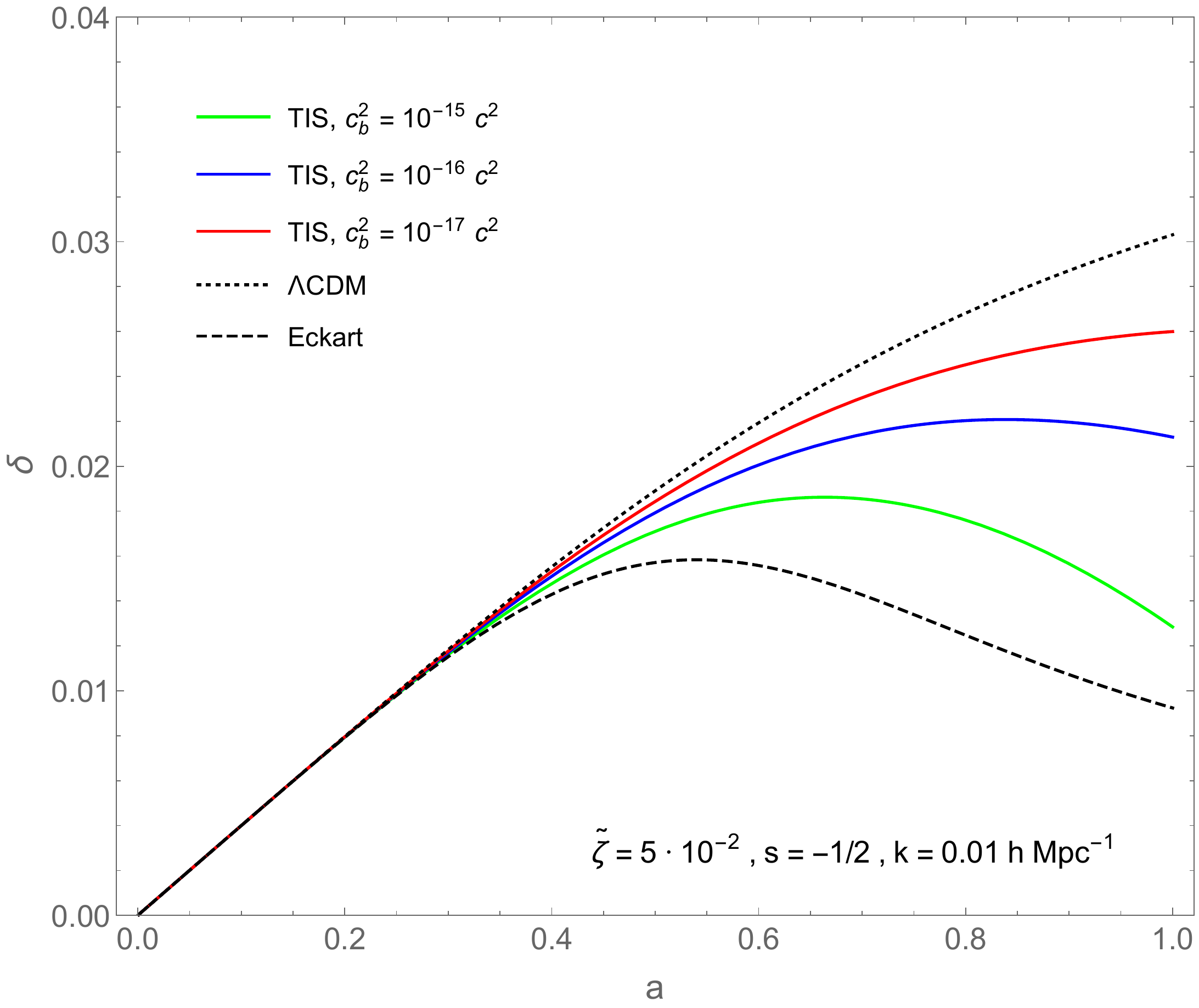}
    \includegraphics[width=0.47\textwidth]{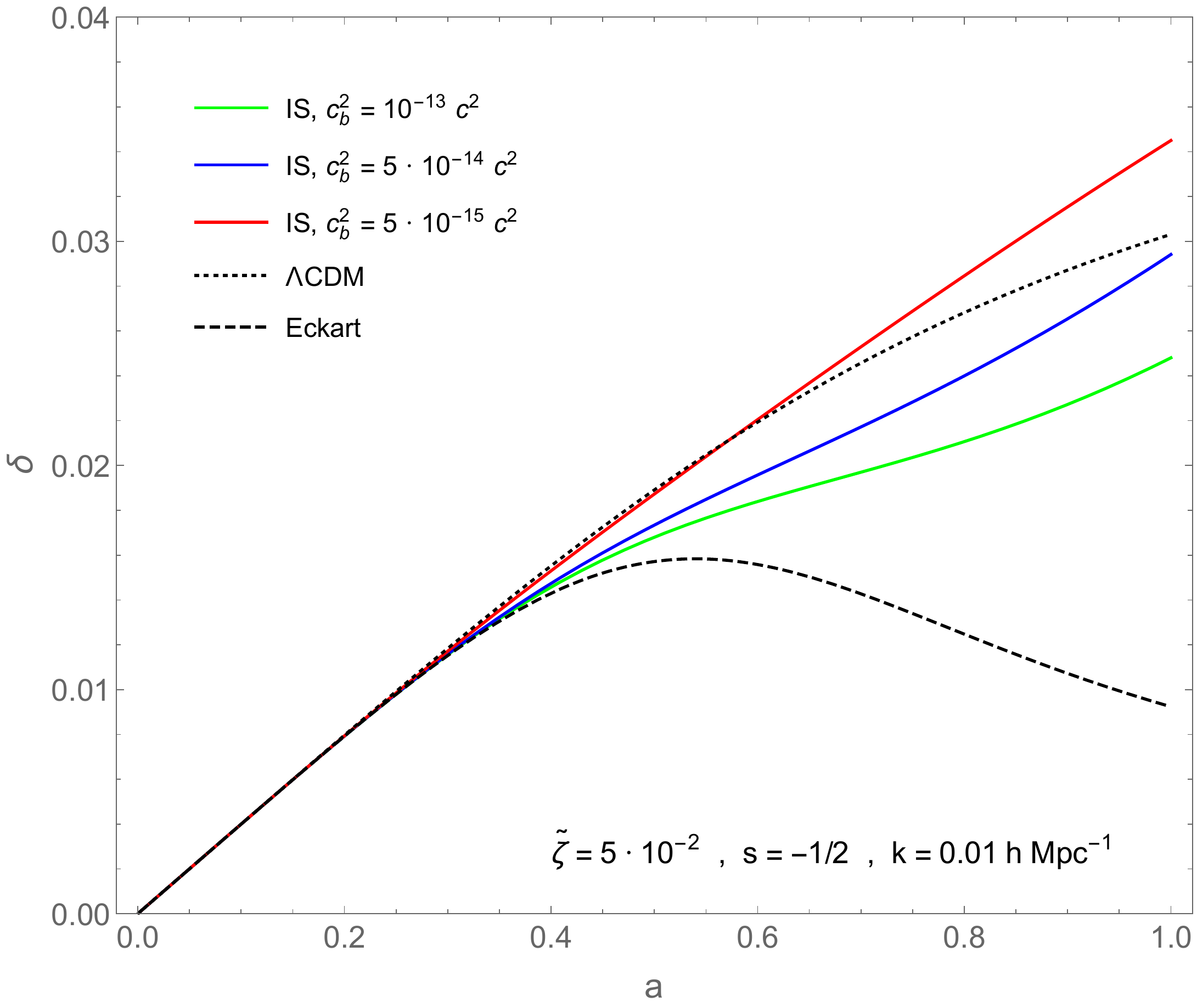}
    \includegraphics[width=0.455\textwidth]{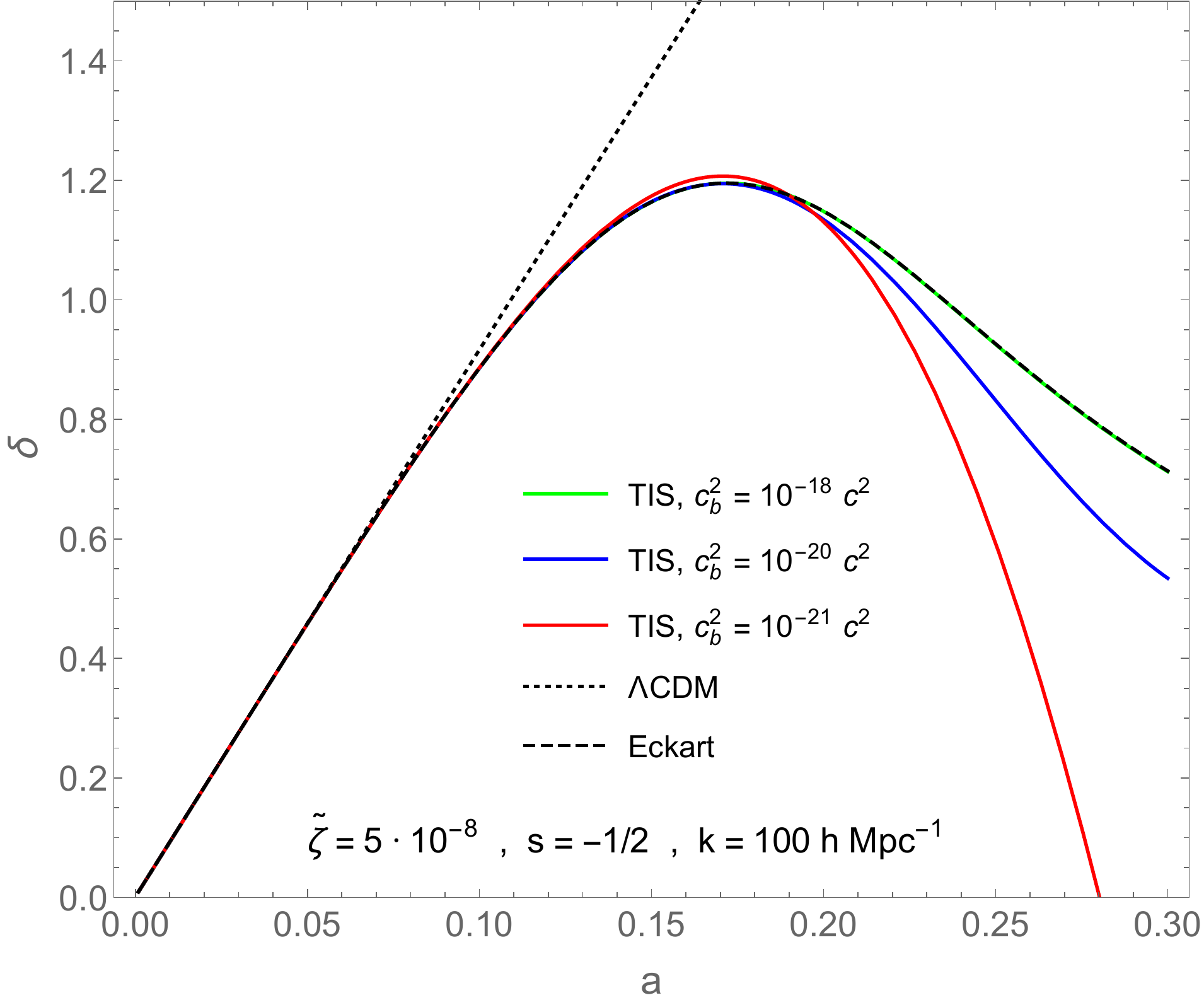}
    \includegraphics[width=0.455\textwidth]{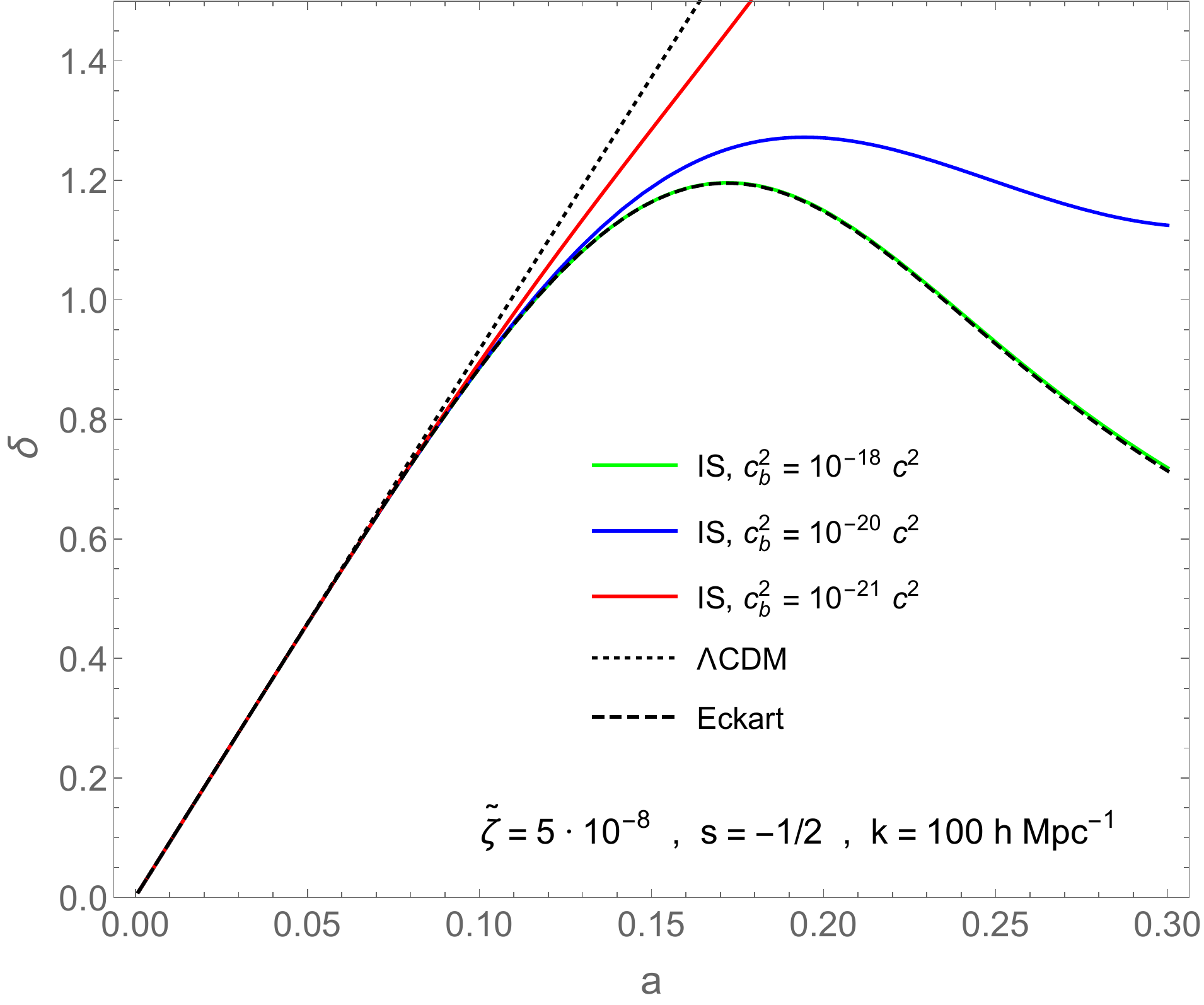}
   \caption{\label{delta_s-05} Evolution of the density contrast for $s=-1/2$ varying $c_b^2$ in TIS (left panels) and IS (right panels), for $k=0.01\ h$ Mpc$^{-1}$ (top panels) and $k=100\ h$ Mpc$^{-1}$ (bottom panels).}
  \end{center}
\end{figure*}

The second case we consider still belongs to the class with $s<1/2$ but with a negative value.  Fig. \ref{delta_s-05} shows the evolution of the density contrast for $s=-1/2$:  as regards the overall behaviour, this scenario shares similarities with $s=0$ in that both exhibit deviations only at late times.  Unlike the previous case, this scenario presents a stronger scale-dependence: at small $k$ both TIS and IS induce an enhancement of density contrast with respect to Eckart, whereas at large $k$ only IS does.

\begin{figure*}[th!!!]
  \begin{center}
    \includegraphics[width=0.47\textwidth]{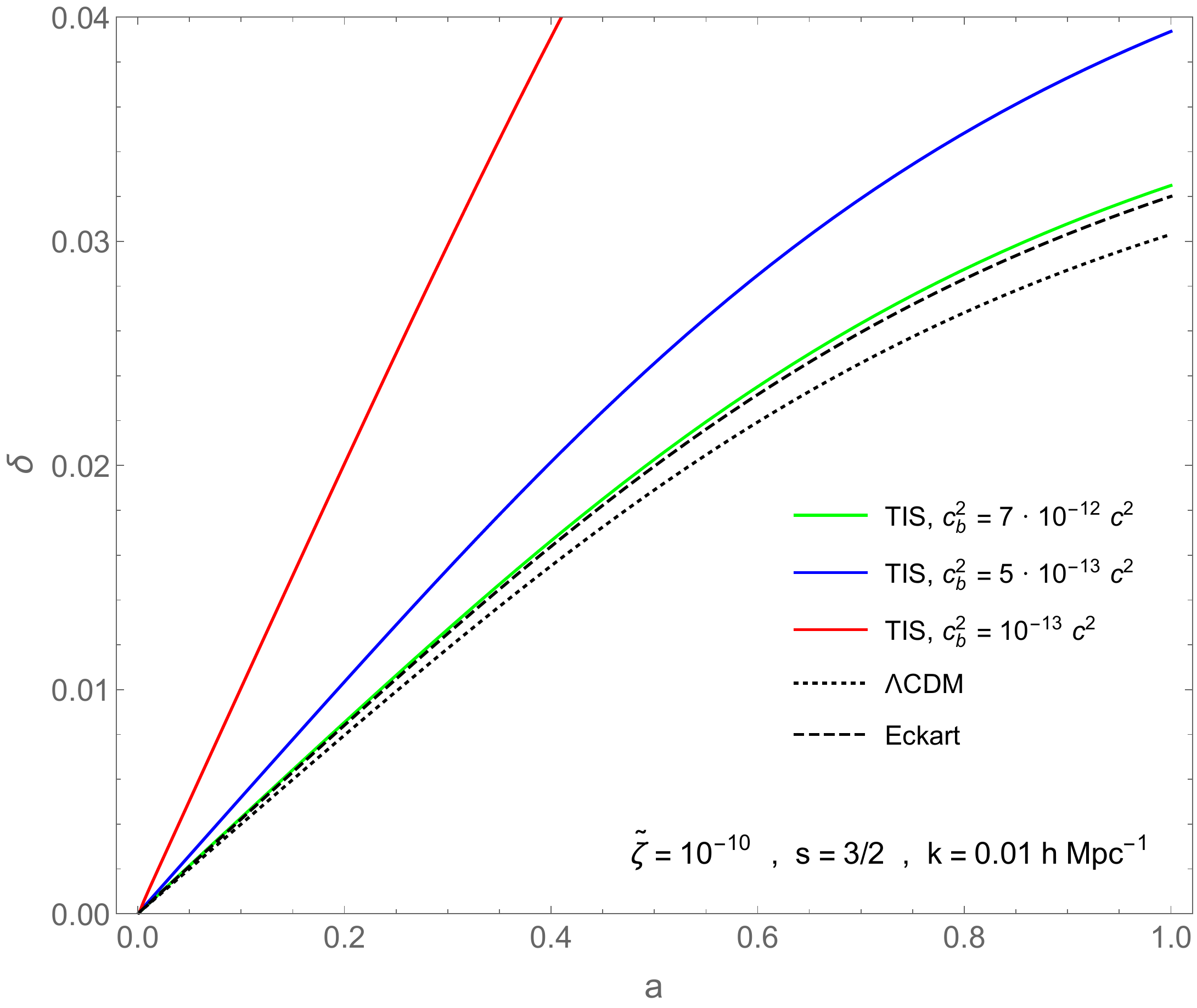}
    \includegraphics[width=0.47\textwidth]{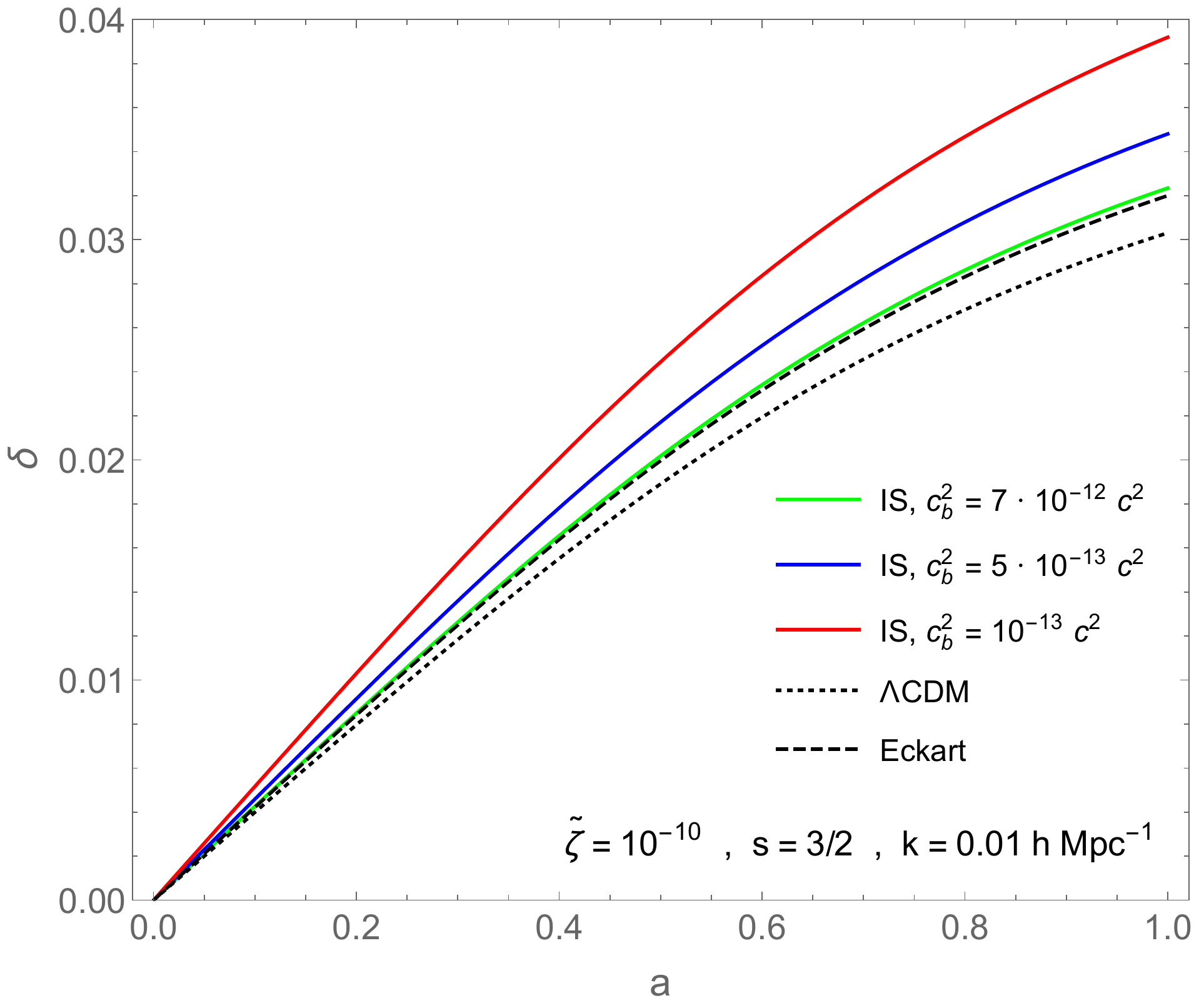}
    \includegraphics[width=0.455\textwidth]{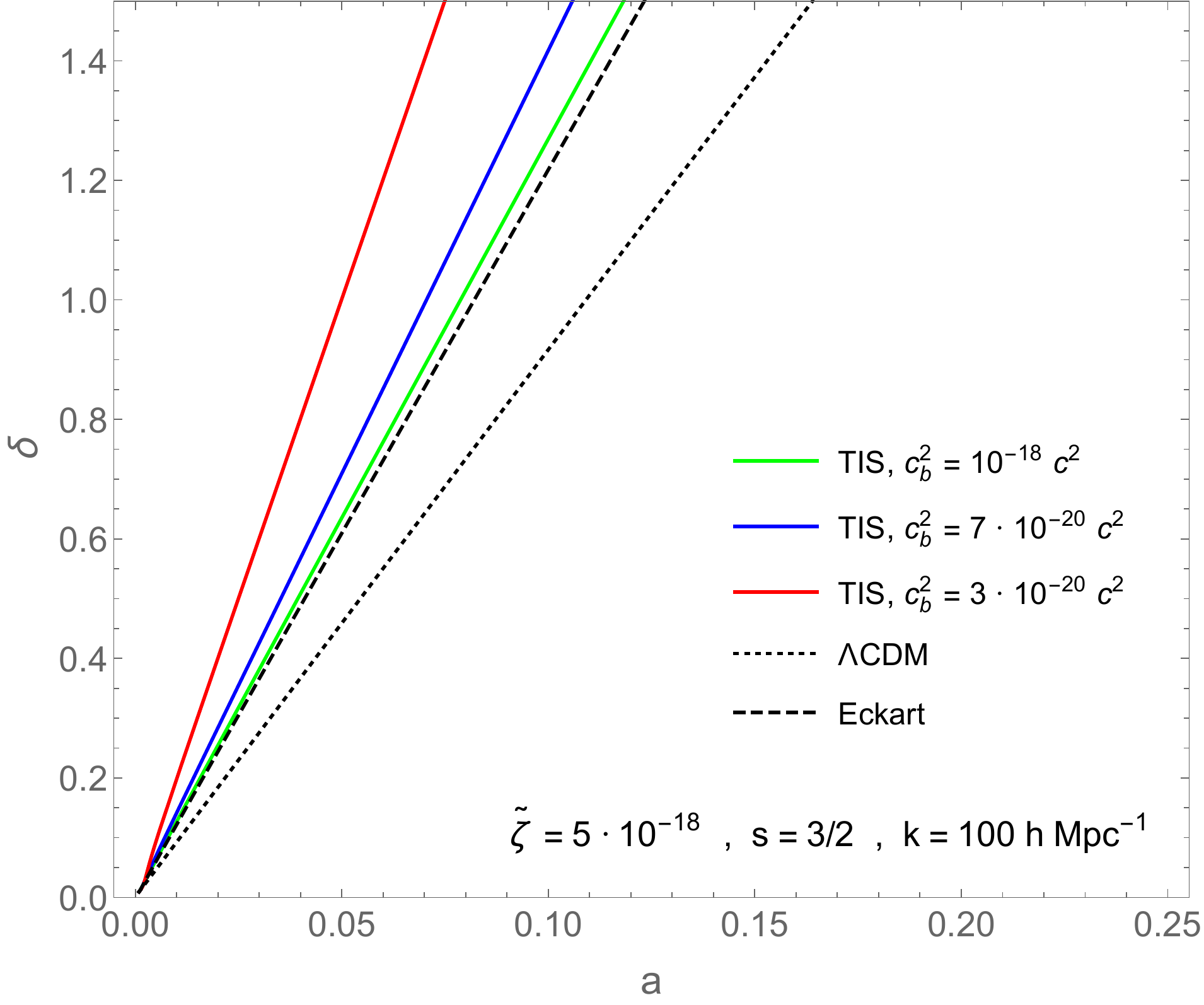}
    \includegraphics[width=0.455\textwidth]{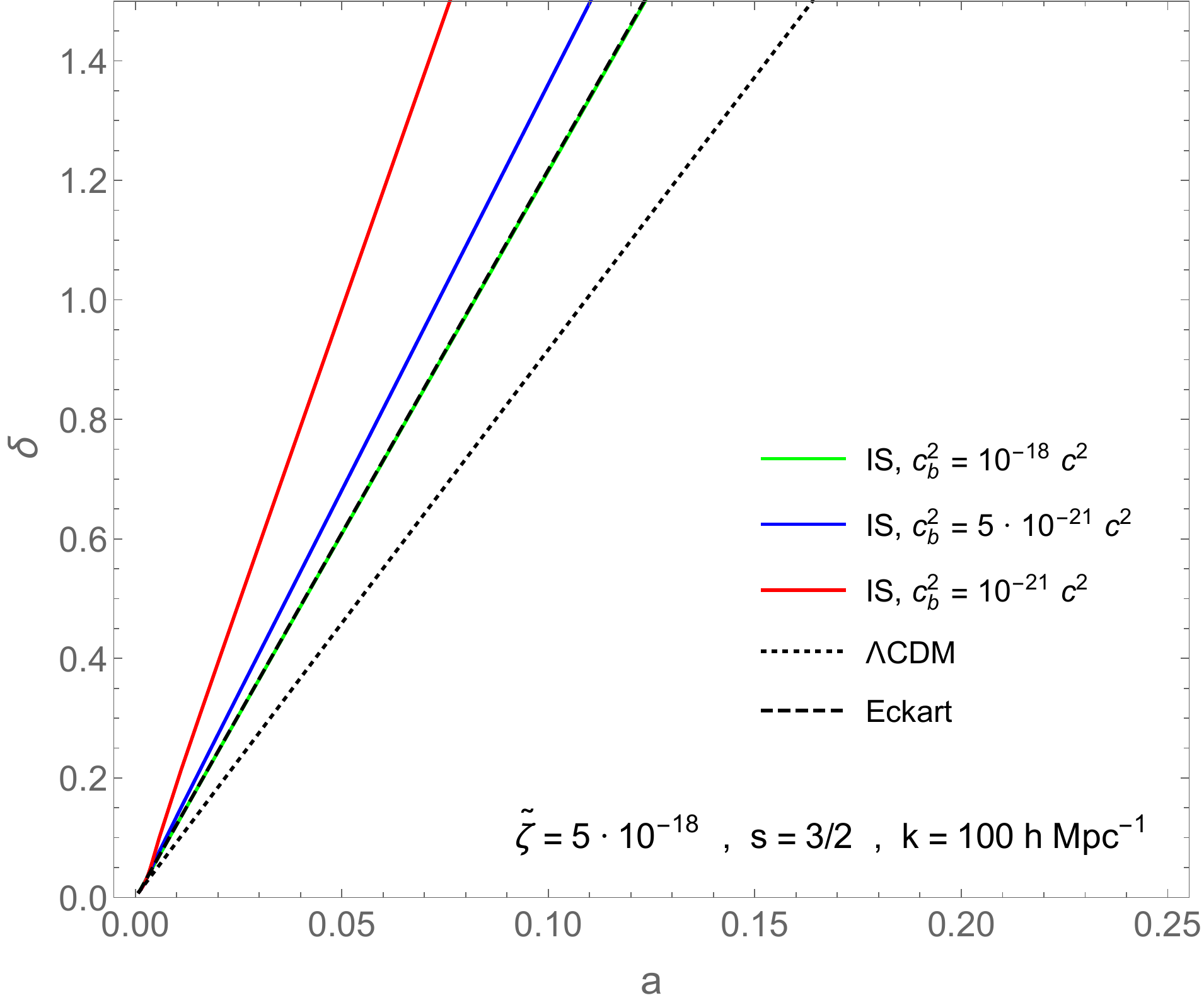}
  \caption{\label{delta_s15} Evolution of the density contrast for $s=3/2$ varying $c_b^2$ in TIS (left panels) and IS (right panels), for $k=0.01\ h$ Mpc$^{-1}$ (top panels) and $k=100\ h$ Mpc$^{-1}$ (bottom panels).}
  \end{center}
\end{figure*}

\begin{figure*}[th!!!]
  \begin{center}
    \includegraphics[width=0.455\textwidth]{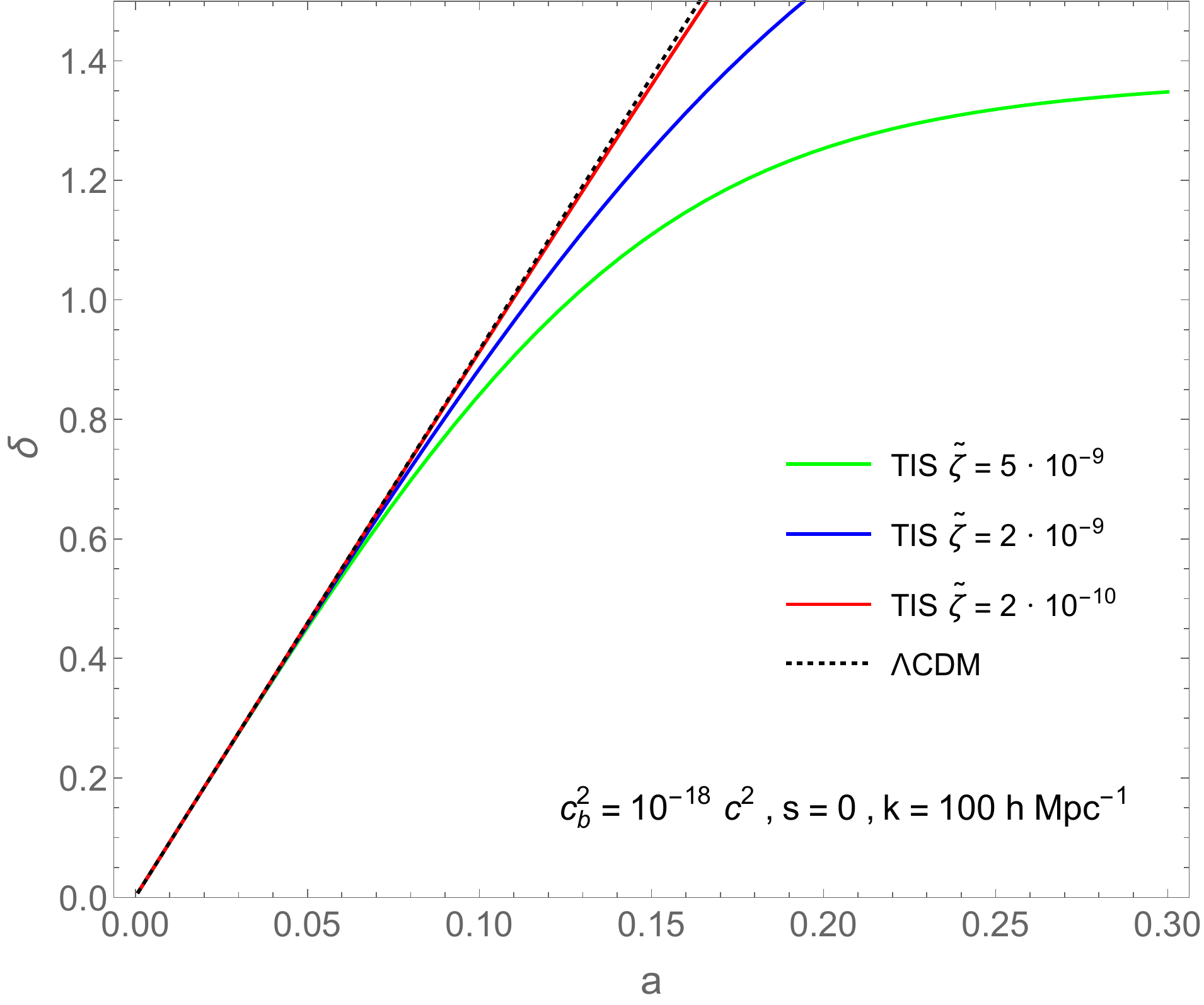}
    \includegraphics[width=0.47\textwidth]{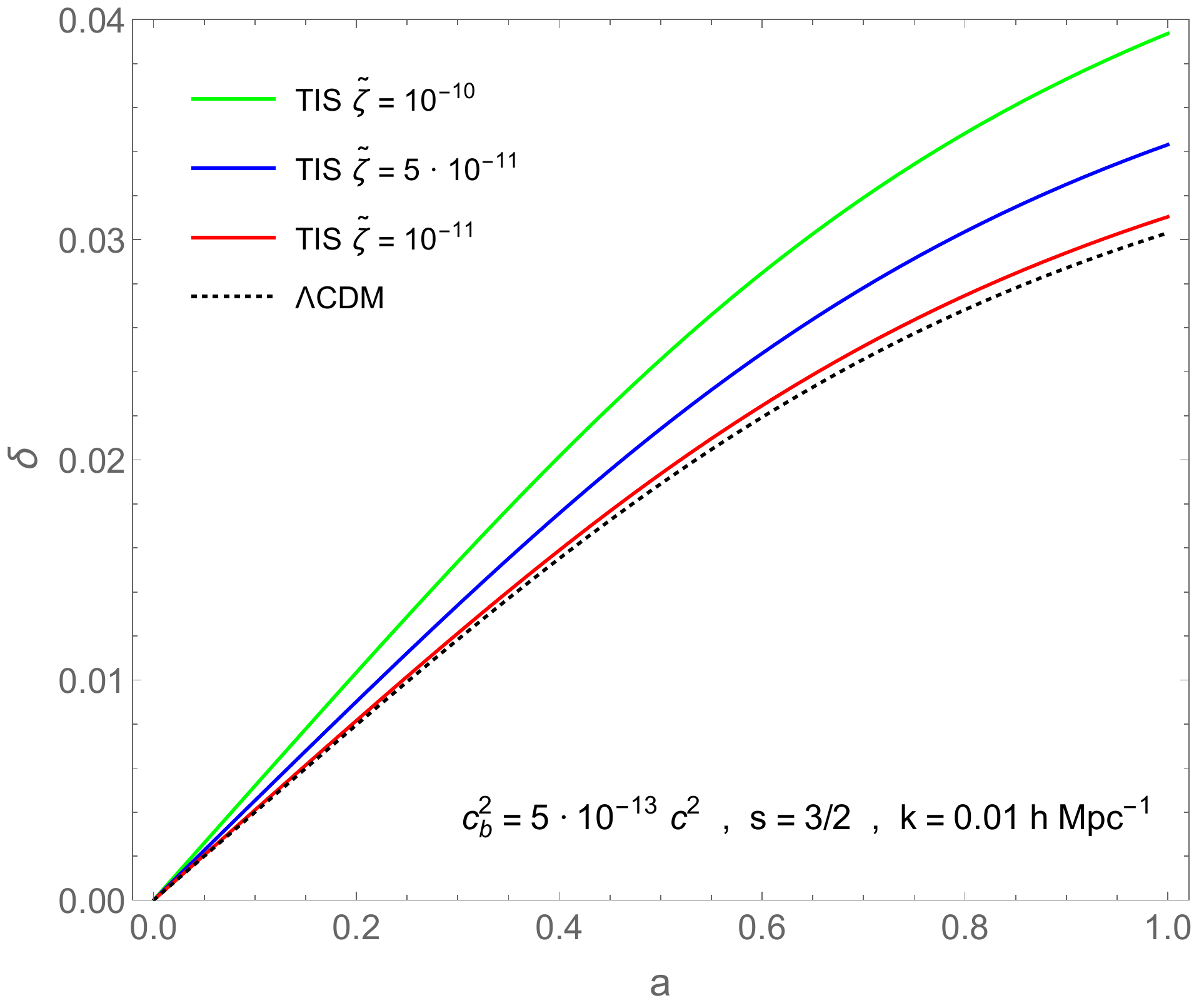}
   \caption{\label{delta_back} Evolution of the density contrast for $s=0$ (left panel) and $s=3/2$ (right panel) varying $\tilde{\zeta}$ in TIS.}
  \end{center}
\end{figure*}

As described in Sect. \ref{qualitative} we expect different features in the perturbative evolution when the exponent of bulk viscosity $s>1/2$.   This is evident when comparing the previous examples with Fig. \ref{delta_s15}, which is for $s=3/2$.  For the latter, all deviations from $\Lambda$CDM show up at earlier times; the Eckart approach leads to an overall enhancement of perturbations with respect to $\Lambda$CDM; both TIS and IS further strengthen this increase.  This is an example of how the deviation from $\Lambda$CDM already present in Eckart is even more pronounced in both TIS and IS at fixed $\tilde{\zeta}$.
\\
Finally, in Fig.\ref{delta_back} we check that in the limit $\tilde{\zeta}\rightarrow 0$ the $\Lambda$CDM scenario is recovered.  This is indeed the case, regardless of scale or the values of $c_b^2$ and $s$. This is a general expected result that holds for Eckart as well as for TIS and IS.

The density contrast can hardly be measured in observations but its derivative, the growth factor $f = d\log\delta/d\log a$, can be obtained in galaxy surveys. In the radial direction, the motion of galaxies depends on the expansion as well as on their peculiar velocities which arise from the density field of the matter in which they are embedded. Therefore, the peculiar velocity field of galaxies $\mathbf{u_p}$ is related to the density constrast through the continuity equation $\dot\delta=-\mathbf{\nabla\cdot u_p}$. It can be expressed in terms of the scale factor and the Hubble parameter, 
\begin{equation}
\mathbf{\nabla\cdot u_p} = - H\, f \, \delta.
\end{equation}
For the $\Lambda$CDM model $f(a)=\Omega_m(a)^\gamma$ with $\gamma = 0.545$.  Any significant departure from this value of $\gamma$ would indicate a deviation from $\Lambda$CDM and it is actively searched for \cite{de2013vimos,raccanelli2013testing}.  Measuring any deviation is one of the main aims of future experiments and galaxy surveys, such as the ESA mission Euclid and the Square Kilometer Array amongst others.

\begin{figure*}[th!!!]
  \begin{center}
    \includegraphics[width=0.47\textwidth]{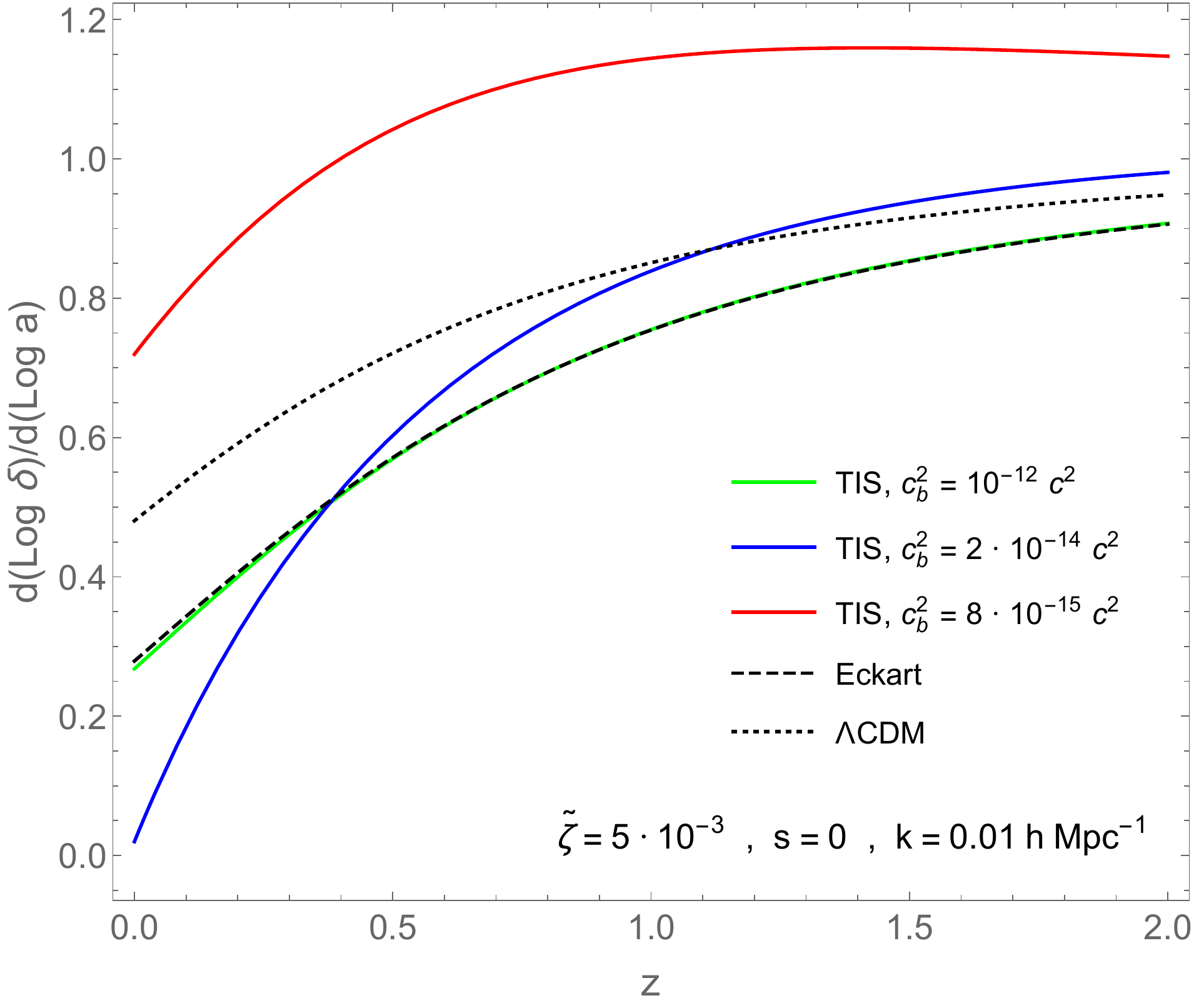}
    \includegraphics[width=0.47\textwidth]{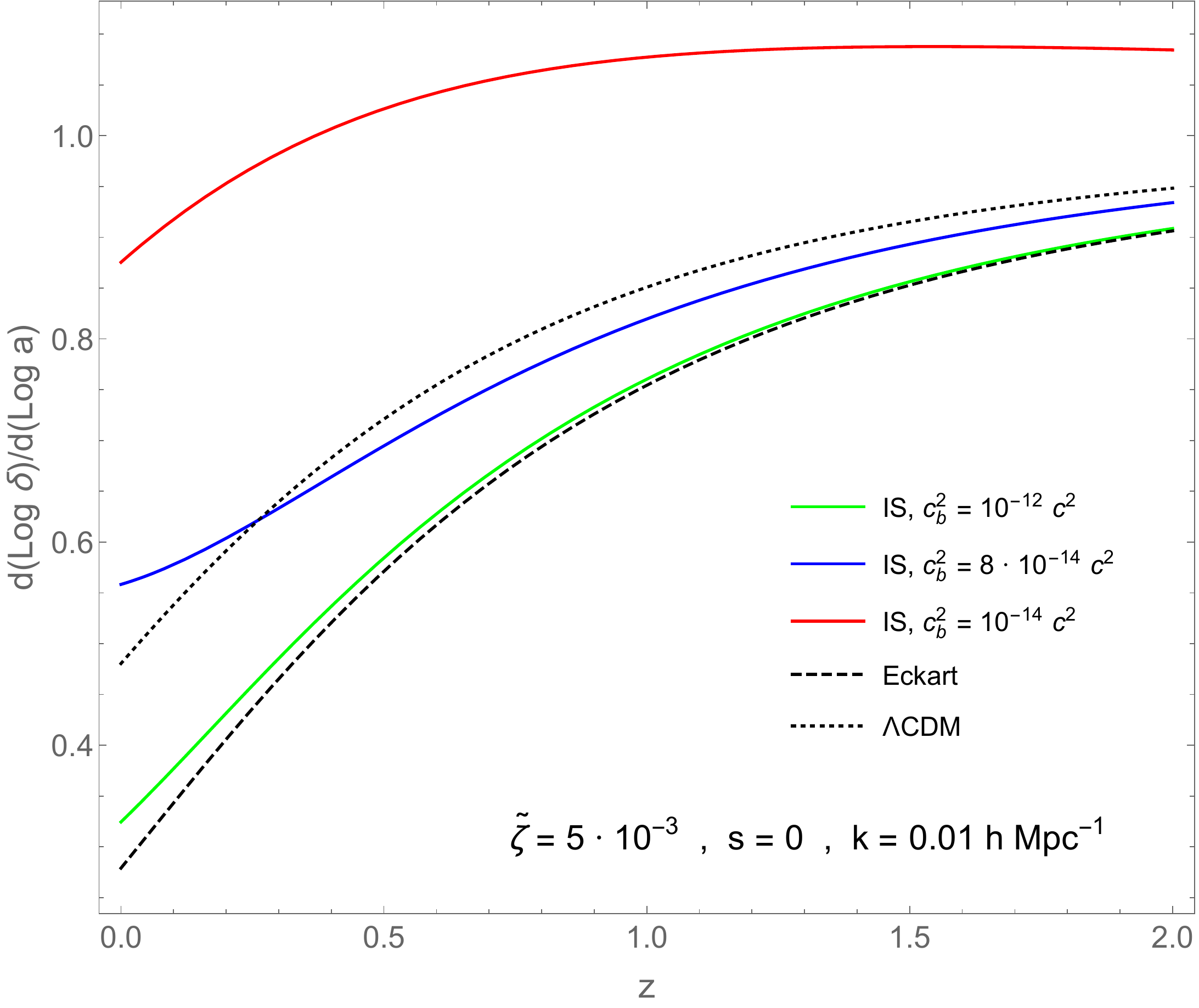}
    \includegraphics[width=0.47\textwidth]{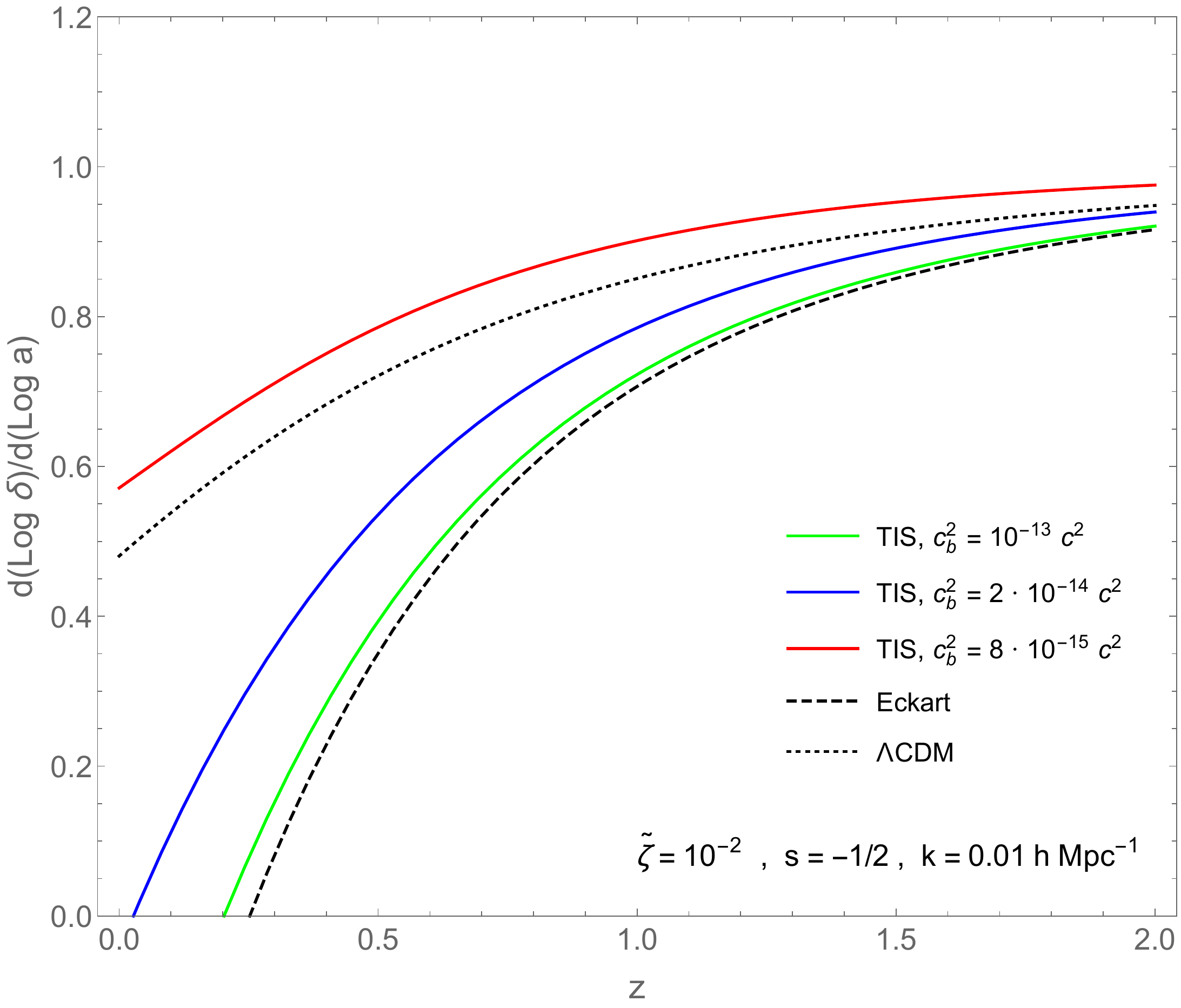}
    \includegraphics[width=0.47\textwidth]{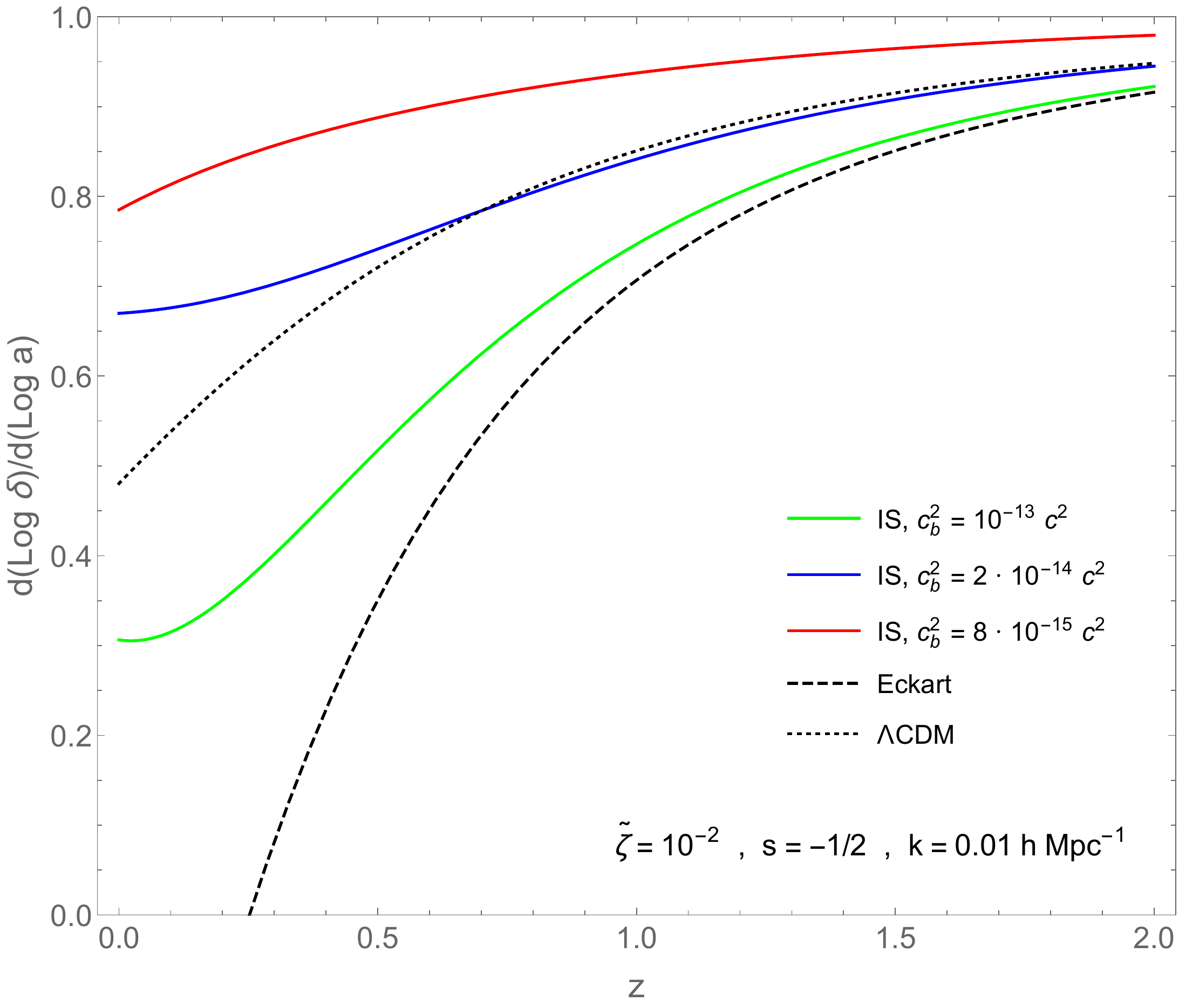}
    \includegraphics[width=0.47\textwidth]{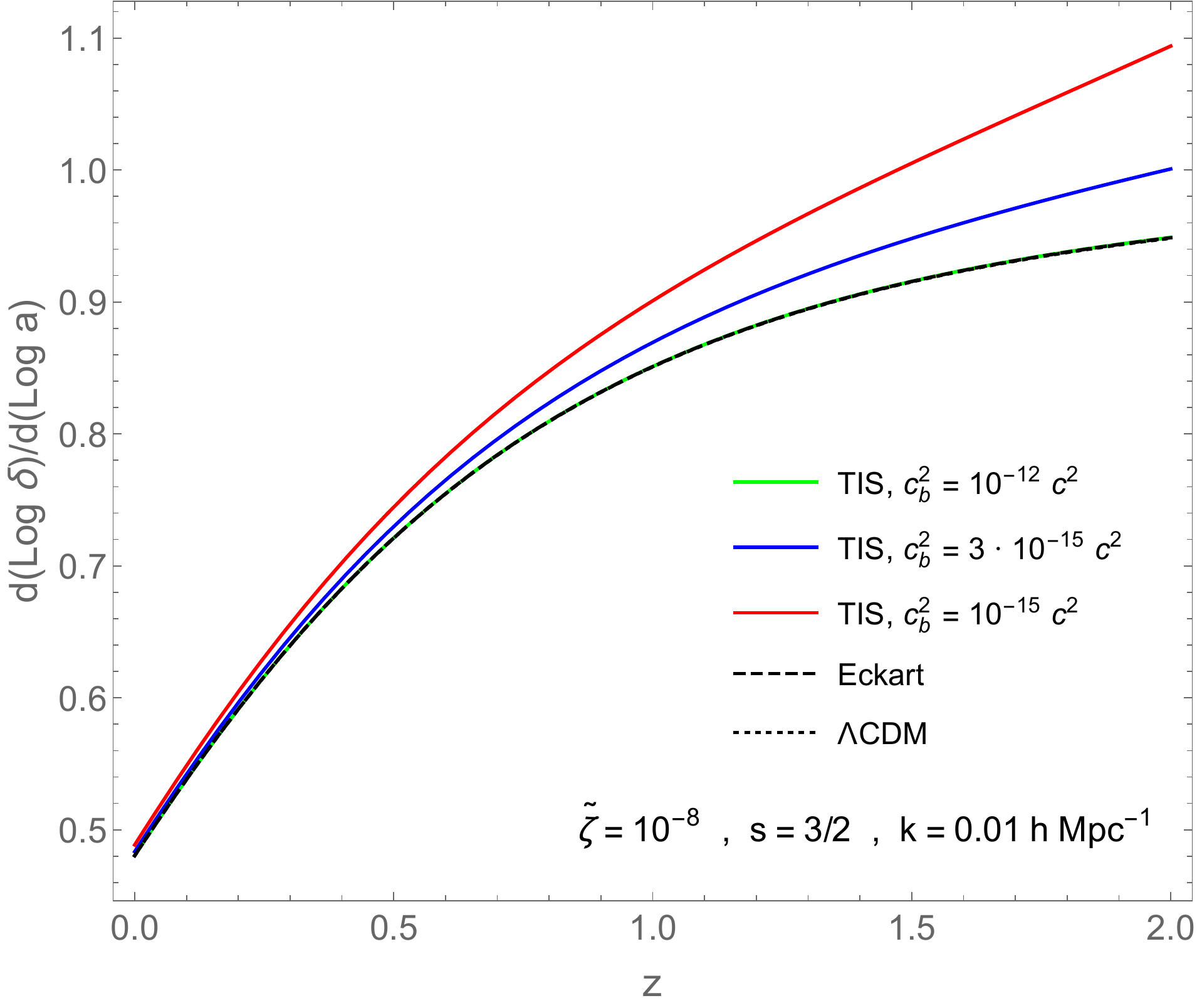}
    \includegraphics[width=0.47\textwidth]{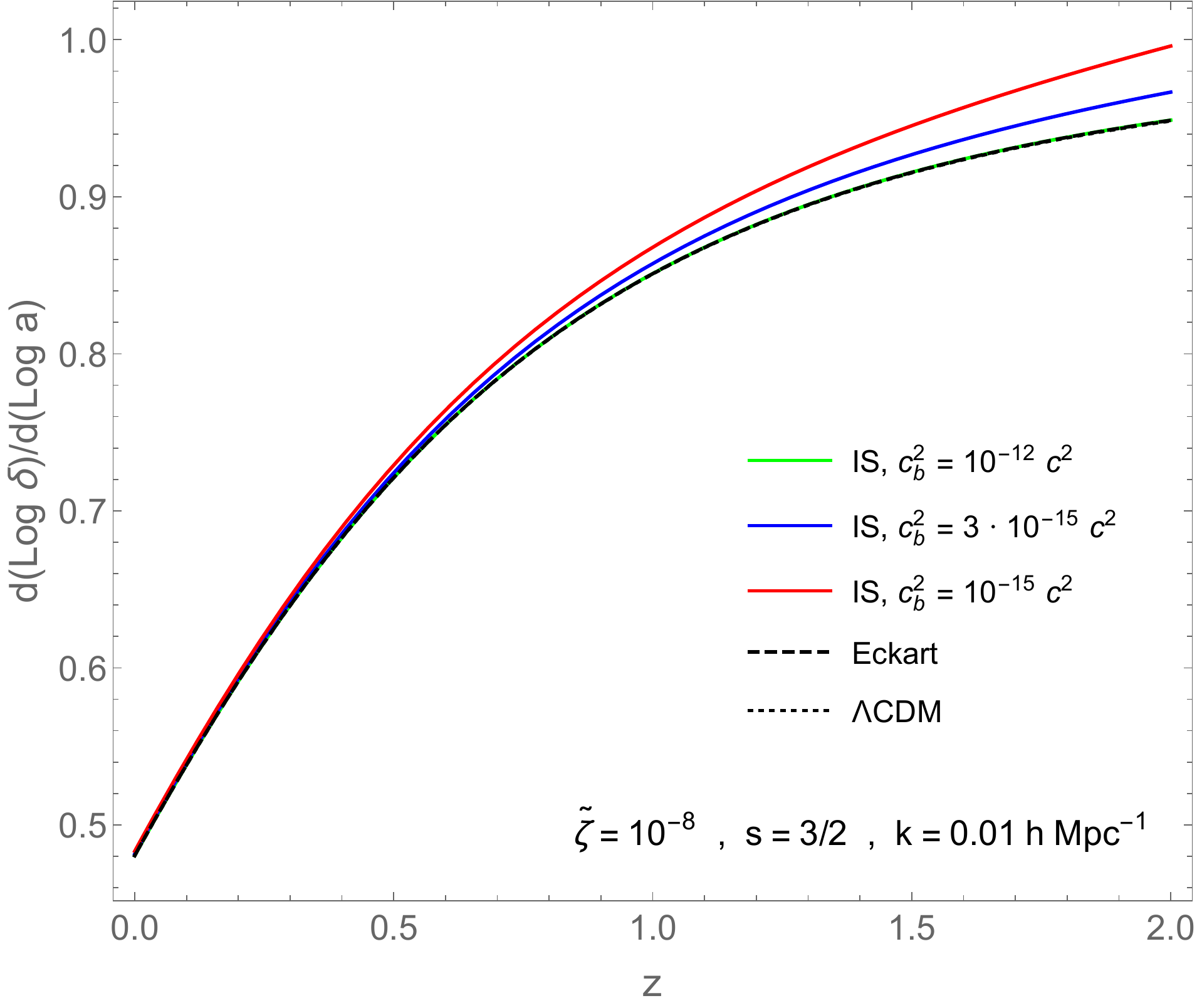}
   \caption{\label{growth_k001} Growth factor as a function of redshift at $k=0.01\, h$ Mpc$^{-1}$, for $s=0$ (top panels), $s=3/2$ (middle panels) and $s=-1/2 $ (bottom panels) in TIS (left panels) and IS (right panels).}
  \end{center}
\end{figure*}

We compare the modifications of the growth factor for $\Lambda$CDM, Eckart, TIS, and IS in Fig. \ref{growth_k001}. For the cases $s=0$, $s=-1/2$ and $s=3/2$ we focus on the scale $k=0.01\, h$Mpc$^{-1}$ over the redshift range $z\in [0,2]$ (which is within reach of present and forthcoming galaxy surveys).

In the $s=0$ case, for both TIS and IS, the variation of $c_b^2$ leads to important modifications in the overall amplitude of the growth factor and to a change of slope at $z\leq 1$. In addition, the amplitude increases as $c_b^2$ decreases.  Interestingly, certain values of $c_b^2$ lead to a growth factor that mimics that of $\Lambda$CDM at $z\geq0.5$ both in the TIS and IS cases.  For $s=-1/2$ the behaviour is qualitatively similar: the deviations are more prominent at small redshifts, while at higher redshifts, if $c_b^2$ is not too small, both TIS and IS are compatible with the $\Lambda$CDM curve.  The similarity with the previous case stems from the fact that both are characterized by $s<1/2$.  In the $s=3/2$ case, which belongs to the $s>1/2$ class, the situation is in fact quite different: the modifications to the $\Lambda$CDM case are of only a few percent for TIS while they are even smaller in the IS case.  Moreover these changes only appear at high redshift.  This is in line with the expectations produced by the qualitative analysis of sec.\ref{qualitative}.


\section{\label{conclusions}Conclusion}

We analysed the effect of causal bulk viscosity on the cosmological dynamics at perturbative level.  In particular, we considered CDM endowed with bulk viscous pressure and  governed by the Israel-Stewart theory of dissipation.  This introduces an additional timescale in the dynamics in the form of a relaxation time $\tau$ that depends on the dissipative sound speed $c_b^2$.  We derived a third order evolution equation for the density contrast and analysed its features both analytically and numerically in several scenarios.  Qualitatively, the deviations from both $\Lambda$CDM and the non-causal Eckart framework are governed by the three free parameters determining the relaxation time: while the ratio between $\zeta_0$ and $c_b^2$ defines the magnitude of the deviation, the value of the exponent $s$ determines whether these deviations occur at early or late times.  In particular we found that $s=1/2$ is a critical value demarcating two scenarios: if $s<1/2$ the deviations show up only at late times, whereas if $s>1/2$ the IS model starts to diverge from the others already at early times. 
 
By considering the truncated version ($\epsilon=0$) instead of the full IS, one essentially eliminates a scale-dependent term in the equations. This has an important influence on the possibility of mimicking the $\Lambda$CDM behaviour.  In the $s<1/2$ cases, the full IS is favoured in this regard, as it tends to evolve inbetween Eckart and $\Lambda$CDM.  In the same cases, the growth factor, even though it can be subject to drastic deviations with respect to $\Lambda$CDM, possesses parameter ranges for which the deviations are small when $z\geq 1$.  The case $s>1/2$ instead presents a different qualitative behaviour: neither TIS nor IS are able to mimic the density contrast of $\Lambda$CDM, but systematically overestimate it. However, the growth factor turns out to be more stable for $\Lambda$CDM, with significant deviations appearing only for $z\geq 1$.  Hence, allowing dark matter to have a bulk viscosity with $s<1/2$ in the context of IS could mitigate the problem of excess of clustering encountered within the framework of $\Lambda$CDM. On the other hand, a choice of $s>1/2$ leads to a further increase of clustering and a greater deviation from $\Lambda$CDM.  In this respect, we note that the enhancement of clustering due to viscosity is a somewhat unexpected result, since bulk viscous effects usually lead to a suppression of the growth of perturbations.  Such behaviour was already found in \cite{Velten2013} for an exponent $s=1/2$ in the Eckart framework, albeit in a non-Newtonian setting.  Given the results of our analysis, we conjecture that such inversion of trend is due to the first-derivative term in Eq.\eqref{ppprime} and that it occurs for some threshold value of the exponent $s$.  However, an analysis of the modes of Eq.\eqref{ppprime} is expected to present a richer structure than in the case of a simple forced/damped oscillator -- as encoutered instead in Eckart's framework.  We plan to obtain more details about this point by analysing a generalized Jeans mechanism that takes into account causal viscosity.

In conclusion, while in Eckart's framework it is possible to approach the $\Lambda$CDM behaviour only by letting $\zeta_0\rightarrow 0$ (vanishing viscosity), with IS it is possible to mimic the standard cosmological model with a nonvanishing viscosity by tuning the additional parameter $c_b^2$.  At the same time, causality (expressed by the very presence of such a parameter) is a physically reasonable requirement for dissipative processes.  This means that IS framework introduces an additional physical degree of freedom to the description, leading to the possibility of describing the cosmic fluids in a less idealised way.  However, we stress that the present analysis relies on a Newtonian approximation. In order to fully explore the effect of causality in viscous dynamics it is necessary to set up a fully relativistic perturbative study, which will be part of future investigations.

\section*{Acknowledgements}
The authors acknowledge useful discussions with Carlo Schimd, Hermano Velten and Dominik Schwarz. We thank the anonymous referee for helpful comments.  Part of this work has been supported by the NITheP short term visitor program.  AP is funded by a NRF SKA Postdoctoral Fellowship. AJ is funded by the NRF Scarce Skills Postdoctoral Fellowship program.  GA is funded by the grant GACR-14-37086G of the Czech Science Foundation.


\appendix

\section{\label{deriv}Derivation of the evolution equation}

To derive the evolution equation for density perturbations, the last term in Eq.\eqref{ddot}
has to be dynamically determined by the perturbed IS equation.  The latter is obtained by perturbing Eq.\eqref{IS} to first order and making use of eqs.\eqref{zeta},\eqref{tau} and \eqref{temp} in eqs.\eqref{dzeta} and
\eqref{dtau}:
\begin{eqnarray}\label{ISpert}
 \tau\, &&\dot{(\delta\Pi)} =\ \frac{\rho a^2}{k^2} \left[ 1+3\, \frac{\epsilon}{2} \left( \frac{2+3w}{1+w} \right) H\, \tau \right]\, \ddot{\delta}\nonumber\\
 &&+ \left\{ \zeta + \frac{\epsilon}{2} \left( \frac{2+3w}{1+w} \right) \Pi\, \tau + \frac{\rho a^2}{k^2} \left[  1+3\, \frac{\epsilon}{2} \left( \frac{2+3w}{1+w} \right) H\, \tau  \right] \right\} \dot{\delta}\nonumber\\
 &&+ \left\{ (s-1)\Pi -3H \zeta - \frac{\rho a^2}{k^2} \left[  1+3\, \frac{\epsilon}{2} \left( \frac{2+3w}{1+w} \right) H\, \tau  \right] \left( 4\pi G\, \rho - w\frac{k^2}{a^2} \right) \right\}\, \delta
\end{eqnarray}
We then differentiate Eq.\eqref{ddot} with respect to time. In the resulting third order equation there will be
$\dot{(\delta\Pi)}$ and $\delta\Pi$ terms, which can be expressed in terms of $\delta$ and its derivatives by
means of Eq.\eqref{ISpert} and Eq.\eqref{ddot}.  As a result, rearranging the terms, one obtains the
following general equation for the evolution of density perturbations:
\begin{align}\label{dddot}
 \tau\, \dddot{\delta} &+ \left\{ \left[ 1 + \frac{3\, \epsilon}{2} \left( \frac{2+3w}{1+w} \right) \right] H \tau + 1 \right\}\, \ddot{\delta}\nonumber\\
 &+ \left\{ \left[ 2\dot{H} - 2H^2 - 4\pi G\rho + w\frac{k^2}{a^2} + \epsilon \left( \frac{2+3w}{1+w} \right)\left( 3H^2+\frac{k^2\, \Pi}{2a^2\rho} \right) \right] \tau + 2H+\frac{k^2 \zeta}{a^2 \rho} \right\}\, \dot{\delta}\nonumber\\
 &+ \left\{ \left[ 16\pi G \rho - 3w\frac{k^2}{a^2}- \frac{3\, \epsilon}{2} \left( \frac{2+3w}{1+w} \right) \left( 4\pi G\rho - w\frac{k^2}{a^2} \right) \right] H \tau \right. \nonumber\\
 &\hspace{1cm}\left. + \frac{k^2}{a^2\rho} \Big( (s-1) \Pi - 3H\zeta \Big) - 4\pi G\rho+w\frac{k^2}{a^2} \right\}\, \delta = 0
\end{align}
The contributions coming directly from the inclusion of causality in the description are the third derivative
$\dddot{\delta}$ and the terms in square brackets.  For dust ($w=0$),
Eq.\eqref{dddot} reduces to
\begin{align}\label{dddotdust}
 \tau\, \dddot{\delta} &+ \Big\{ \left[ 1 + 3 \epsilon \right] H \tau + 1 \Big\}\, \ddot{\delta} + \left\{ \left[ 2\dot{H} - 2H^2 - 4\pi G\rho + 2 \epsilon \left( 3H^2+\frac{k^2\, \Pi}{2a^2\rho} \right) \right] \tau + 2H+\frac{k^2 \zeta}{a^2 \rho} \right\}\, \dot{\delta}\nonumber\\
 &+ \left\{ 4\pi G \rho \left( 4-3\epsilon \right)\, H \tau + \frac{k^2}{a^2\rho} \Big( (s-1) \Pi - 3H\zeta \Big) - 4\pi G\rho \right\}\, \delta = 0
\end{align}
The final form Eq.\eqref{ppprime} used in the present analysis can be obtained from Eq.\eqref{dddotdust} by
defining a new time derivative $\delta'=d\delta/da$, such that $\dot{\delta}=\dot{a}\delta'$.  One can then
easily check that the non-causal Eckart theory is recovered for $\tau\rightarrow 0$ (keeping in
mind that in this case $\Pi=-3H\, \zeta$). Non--viscous perturbations are recovered for $\zeta\rightarrow
0$.


\bibliographystyle{apsrev4-1}
\bibliography{biblio}

\end{document}